\documentclass{vldb}

\makeatletter
\newif\if@restonecol
\makeatother

\usepackage{url}
\usepackage{times}
\usepackage[vlined,algoruled]{algorithm2e}
\usepackage{verbatim}   
\usepackage{color}      
\usepackage{bm}                     	
\usepackage{enumitem}

\usepackage{caption}
\usepackage{subcaption}
\usepackage{balance}
\usepackage{multirow}

\linesnumbered
\SetKwInOut{Input}{Input}
\SetKwInOut{Output}{Output}
\SetKw{Break}{break}
\SetKw{Continue}{continue}
\SetCommentSty{textsl}

\newcommand{\eat}[1]{}
\newcommand{\tightlist}{\itemsep=-2pt}
\newcommand{\rbox}{\hfill $\Box$}
\newcommand{\ie}{{\em i.e.}}
\newcommand{\eg}{{\em e.g.}}

\begin{document}

\numberofauthors{1}
\author{
\alignauthor
Xin Luna Dong, Evgeniy Gabrilovich, Geremy Heitz, Wilko Horn,\\ 
Kevin Murphy, Shaohua Sun, Wei Zhang\\
 \affaddr{Google Inc.}
 \email{\{lunadong|gabr|geremy|wilko|kpmurphy|sunsh|weizh\}@google.com}
}

\title{From Data Fusion to Knowledge Fusion}
\maketitle

\begin{abstract}
{\small
The task of {\em data fusion} is to identify the true values of data items
(\eg, the true date of birth for {\em Tom Cruise})
 among multiple observed values drawn from different sources (\eg, Web
 sites) of varying (and unknown) reliability.
 A recent survey~\cite{LDL+12} has provided a detailed comparison of
 various fusion methods on Deep Web data.
 In this paper, we study the applicability and limitations
 of different fusion techniques on a more challenging problem:
{\em knowledge fusion}. Knowledge fusion identifies
 true subject-predicate-object triples extracted by
 multiple information extractors from multiple information sources.
These extractors perform the tasks of entity linkage and
schema alignment, thus introducing an additional source of noise that is quite
different from that traditionally considered in the data fusion
literature, which only focuses on factual errors in the original sources.
We adapt state-of-the-art data fusion techniques and apply them to
a knowledge base with 1.6B unique knowledge triples extracted
by 12 extractors from over 1B Web pages,
which is three orders of magnitude larger than the data sets
used in previous data fusion papers.
We show great promise of the data fusion approaches in solving the knowledge fusion problem,
and suggest interesting research directions through a detailed error analysis of the methods.
}
\end{abstract}

\newtheorem{definition}{Definition}[section]
\newtheorem{proposition}[definition]{Proposition}
\newtheorem{lemma}[definition]{Lemma}
\newtheorem{remark}[definition]{Remark}
\newtheorem{corollary}[definition]{Corollary}
\newtheorem{claim}[definition]{Claim}
\newtheorem{theorem}[definition]{Theorem}
\newtheorem{example}[definition]{Example}

\newtheorem{review}{Comment}[subsection]
\newcommand{\answer}[1]{{\bf Answer:} #1}

\section{Introduction}
\label{sec:intro}
Extracting information from multiple, possibly conflicting, data sources, and reconciling
the values so the true values can be stored in a central data repository, is a problem of
 vital importance to the database and knowledge management communities.
A common way to formalize the problem is to assume we have $M$ data
items, each describing a particular aspect of an entity
(\eg, the date of birth of {\em Tom Cruise}),
and $N$ data sources (\eg, Web sites); we can visualize the raw data as an $M
\times N$ data matrix, where each cell represents the values provided by a source
on a data item, although many cells may be empty, representing missing data
(see Figure~\ref{fig:input}(a)).
The goal is to infer the true latent value for each of
the data items (rows), given the noisy observed values in the matrix,
while simultaneously estimating the unknown quality and interdependence of
the data sources (columns). This problem is called {\em data fusion},
and has received a lot of attention in recent years
(see~\cite{BN08, LDL+12} for recent surveys).

In this paper, we study how we can adapt existing data fusion techniques
to the more challenging area of
automatically constructing large-scale knowledge bases (KBs).
To build a knowledge base,
we employ multiple knowledge extractors to extract (possibly conflicting)
values from each data source for each data item; we then need to
decide the degree of correctness
of the extracted knowledge.
We call this new problem {\em knowledge fusion}.
We can visualize it as adding a third dimension
to the aforementioned data matrix (see Figure~\ref{fig:input}(b)).
This third dimension corresponds to different information extractors applied to
each data source. These extractors must convert raw data, often
expressed in unstructured text format, into structured knowledge that
can be integrated into a knowledge base. This involves three key steps:
identifying which parts of the data indicate a data item and its value;
linking any entities that are mentioned to the
corresponding entity identifier; and linking any relations that are
mentioned to the corresponding knowledge base schema. All three steps are error prone.
Furthermore, the errors introduced by these steps are quite different
from what has been considered in data fusion, which has focussed on factual errors
in the original sources. To handle the increased level of noise in the
data, we compute a calibrated probability distribution over values for
each data item, rather than just returning a single ``best guess'' as
in data fusion.

We make three contributions in this paper.
The first contribution is to define the knowledge fusion problem,
and to adapt existing data fusion techniques to
solve this problem.
We select three such techniques and devise efficient MapReduce
based implementations for them.
We evaluate the performance of these techniques on a knowledge base that contains
 1.6B RDF triples in the form of ({\tt subject}, {\tt predicate}, {\tt object}),
extracted by 12 extractors from over 1 billion Web pages;
the size of the data is 1000 times larger than any data set used
in previous data fusion experiments.

The second contribution is to suggest some simple improvements to
existing methods that substantially improve their quality, especially
in terms of the quality of the calibration of the estimated
probabilities.
The result of these improvements is a fairly well calibrated system:
when we predict a probability above 0.9, the real accuracy of the triples
is indeed high (0.94); when we predict a probability below 0.1,
the real accuracy is fairly low (0.2); and when we predict
a medium probability in [0.4, 0.6), the real accuracy also matches
well (0.6).

The third contribution is a detailed error analysis of our methods,
and a list of suggested directions for future research to
address some of the new problems raised by knowledge fusion.

\smallskip
\noindent
{\bf Related work:}
There are two main bodies of related work. The first concerns data
fusion, surveyed in \cite{LDL+12}. As discussed above, the
current paper adapts those approaches to solve the more challenging
problem of knowledge fusion by considering a third source
of noise from the use of multiple information extractors.
The second main body of work is related to knowledge base construction,
such as 
YAGO~\cite{yago}, 
NELL \cite{CBK+10}, and DeepDive \cite{Niu12}.
Prior work in this literature focuses on
applying (semi-)supervised machine learning methods to improve
 extraction quality, while we focus on resolving conflicts from different
systems using an unsupervised approach, treating the extractors
themselves as black boxes.

Finally, we note the difference of our work from the domain-centric
approach that extracts and integrates structured data in a particular
domain. Dalvi et al.~\cite{DMP12} studied distribution of structured data
in particular domains to evaluate the feasibility and efficacy
of such an approach, whereas we focus on detecting errors from
knowledge extracted from various types of data in domains of a large variety.

\smallskip
The rest of the paper is structured as follows.
Section~\ref{sec:fusion} briefly surveys the state-of-the-art methods in data fusion.
Section~\ref{sec:kv} describes knowledge extraction and formally defines
knowledge fusion. Section~\ref{sec:performance} evaluates the performance
of data fusion techniques on knowledge fusion and presents
a detailed error analysis. Section~\ref{sec:improvement} suggests future
research directions and Section~\ref{sec:conclude} concludes.

\section{State of the Art in Data Fusion}
\label{sec:fusion}
\begin{figure}
\vspace{-.2in}
\hspace{-.4in}
\includegraphics[scale=0.4]{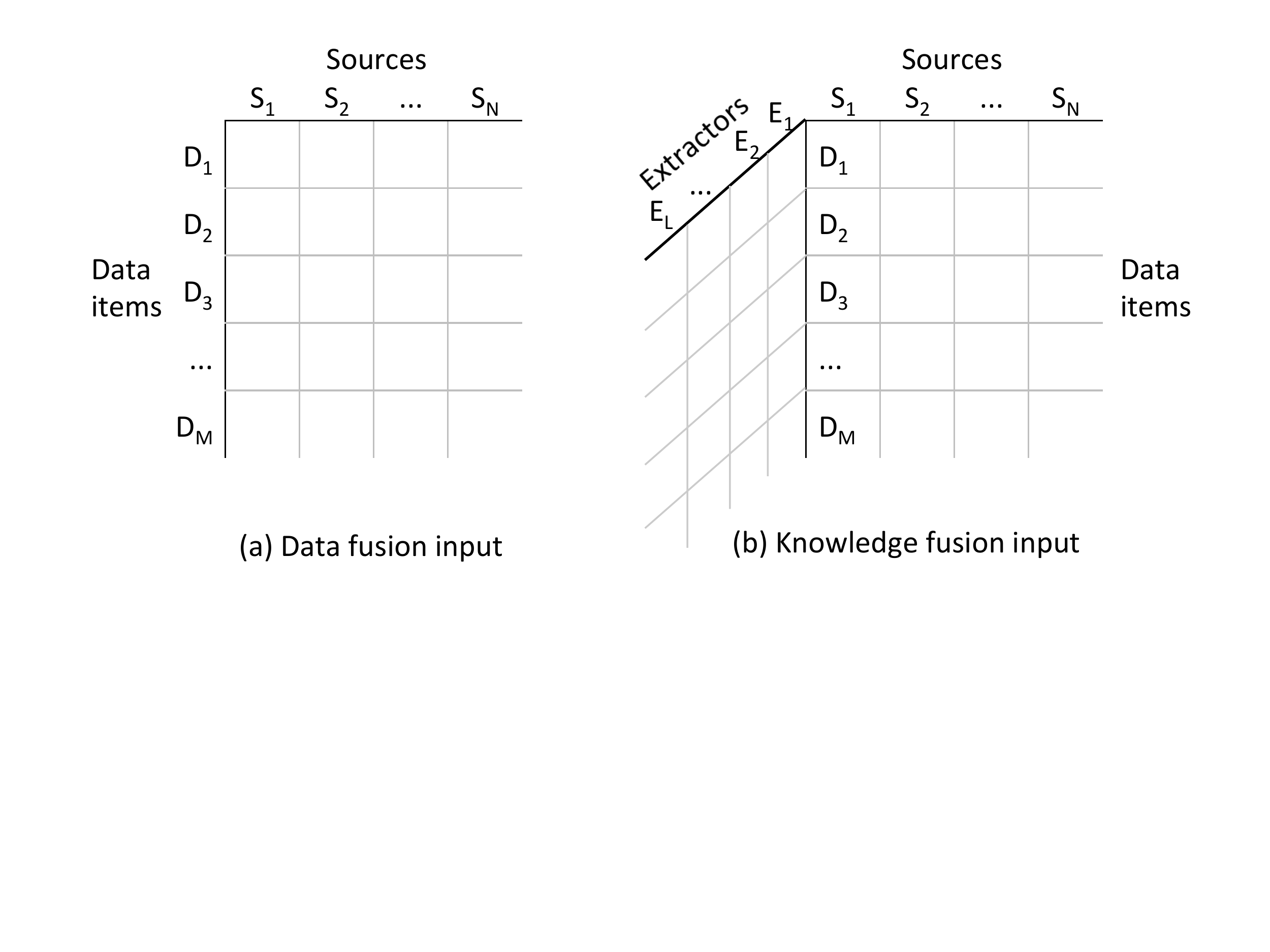}
\vspace{-1.4in}
\caption{Input for data fusion is two-dimensional
whereas input for knowledge fusion is three-dimensional.
\label{fig:input}}
\vspace{-.1in}
\end{figure}

Data fusion is the problem of resolving conflicting values
from different sources, and finding the underlying true values.
In this section, we provide a brief summary of existing approaches to
this problem; this will set the stage for our later discussion of
knowledge fusion.

We can consider the input of data fusion as a two-dimensional data matrix (Figure~\ref{fig:input}(a)).
Each row represents a data item that describes
a particular aspect of an entity, such as the {\em birth date} of {\em Tom Cruise}.
Each column represents a data source.
Each cell represents the value (or values) provided by the corresponding source
on the corresponding data item. The matrix can be sparse, since a source may provide
values on only a subset of the data items. Erroneous values may be provided
and conflicts may exist for the same data item.
Data fusion aims at finding the one true value (or
sometimes the set of true values) for each data item.
Note that for this paper we consider only a snapshot of data rather
than temporally evolving
data from data sources that are frequently updated.

Early approaches to data fusion methods were typically rule-based,
such as using the observed value from the most recently updated source,
or taking the average/maximum/minimum for numerical values.
They focus on improving efficiency with the use of database queries
(surveyed in~\cite{BN08, DN09}). Recently many advanced solutions
have been proposed that apply unsupervised learning or semi-supervised learning
to find the truths (see~\cite{LDL+12} for a recent survey and~\cite{PR13, QAH+13}
for works thereafter).
We can roughly classify these methods into three classes.

\smallskip
\noindent
{\bf Voting:} {\em Voting} is a baseline strategy. Among conflicting values,
each value has one vote from each data source, and we take the value with
the highest vote count (\ie, the value provided by the largest number of sources).

\smallskip
\noindent
{\bf Quality-based:} {\em Quality-based methods} evaluate the trustworthiness
of data sources and accordingly compute a higher vote count for
a high-quality source.
We can further divide them into four sub-categories according to how they
measure source trustworthiness. 

\begin{itemize}\tightlist
  \item {\em Web-link based methods~\cite{Kleinberg98, PR10, PR11, YT11}} measure
source trustworthiness and value correctness using  {\em PageRank~\cite{pagerank}}:
a source providing a value is considered as a link between the source and the value.

  \item {\em IR-based methods~\cite{GAMS10}} measure source trustworthiness
as the {\em similarity} between the provided values and the true values.
They use similarity metrics that are widely accepted in information retrieval,
such as cosine similarity. Value correctness is decided by the accumulated
source trustworthiness.

  \item {\em Bayesian methods~\cite{DBS09a, DSS13, YHY07}} measure the trustworthiness of a source by
its {\em accuracy}, which essentially indicates the probability of each of its values being true.
They apply Bayesian analysis to compute the {\em maximum a
  posteriori} or MAP  value for each data item.

  \item {\em Graphical-model methods~\cite{PR13, ZH12, ZRHG12}} apply probabilistic graphical models
to jointly reason about source trustworthiness and value correctness.
\end{itemize}

\noindent
{\bf Relation-based:} {\em Relation-based methods} extend quality-based
methods by additionally considering the relationships between the
sources.
The relationship can be that of copying between a pair of
sources~\cite{BCM+10, DBH+10a, DBS09a, DBS09b, LDOS11};
in this case,
a copier has a discounted vote count for its copied values.
The relationship can also be correlation
among a subset of sources~\cite{PDD+14, QAH+13};
sources in such a subset may be 
considered as one in trustworthiness evaluation.

\smallskip
Most of the advanced methods take an iterative approach to truth finding,
quality evaluation, and relationship detection when applicable.
They do not use training data and are unsupervised.
A few of them assume the existence of labeled data,
which can be used to estimate source accuracy without the need for iterative
algorithms~\cite{DSS13}, or can be used in conjunction with unlabeled data
for optimization purpose~\cite{YT11}; such methods are semi-supervised.

\eat{
To the best of our knowledge, these methods all make a binary decision on whether
a provided value is true or false. However, since they often attach a vote count
or a correctness score to each value, we may transform them into a probability for
a value to be true. For some methods such transformation is easy since the correctness
score in itself is the {\em a posteriori} probability~\cite{YHY07, DBS09a, DSS13}
for the truthfulness of the value. For some other methods, we may normalize the scores
into $[0,1]$.
}

\section{Fusing Extracted Knowledge}
\label{sec:kv}
\begin{figure*}
\vspace{-.1in}
\begin{minipage}[th]{.32\linewidth}
\centering
\vspace{-.35in}
\includegraphics[scale=0.25]{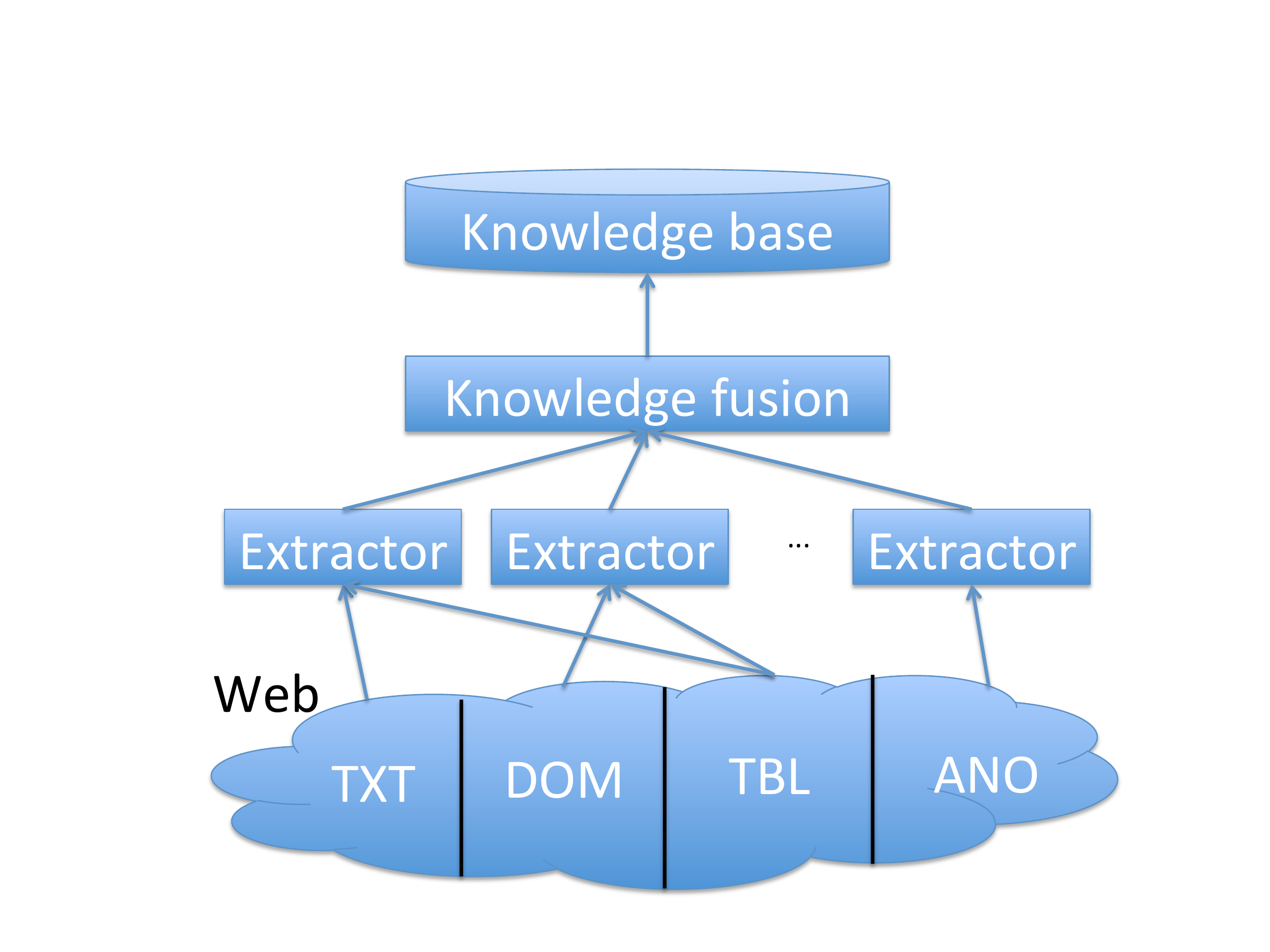}
\vspace{-.2in}
\caption{Architecture of knowledge extraction and fusion.
\label{fig:arch}}
\end{minipage}
\hfill
\begin{minipage}[th]{.32\linewidth}
\centering
\includegraphics[scale=0.24]{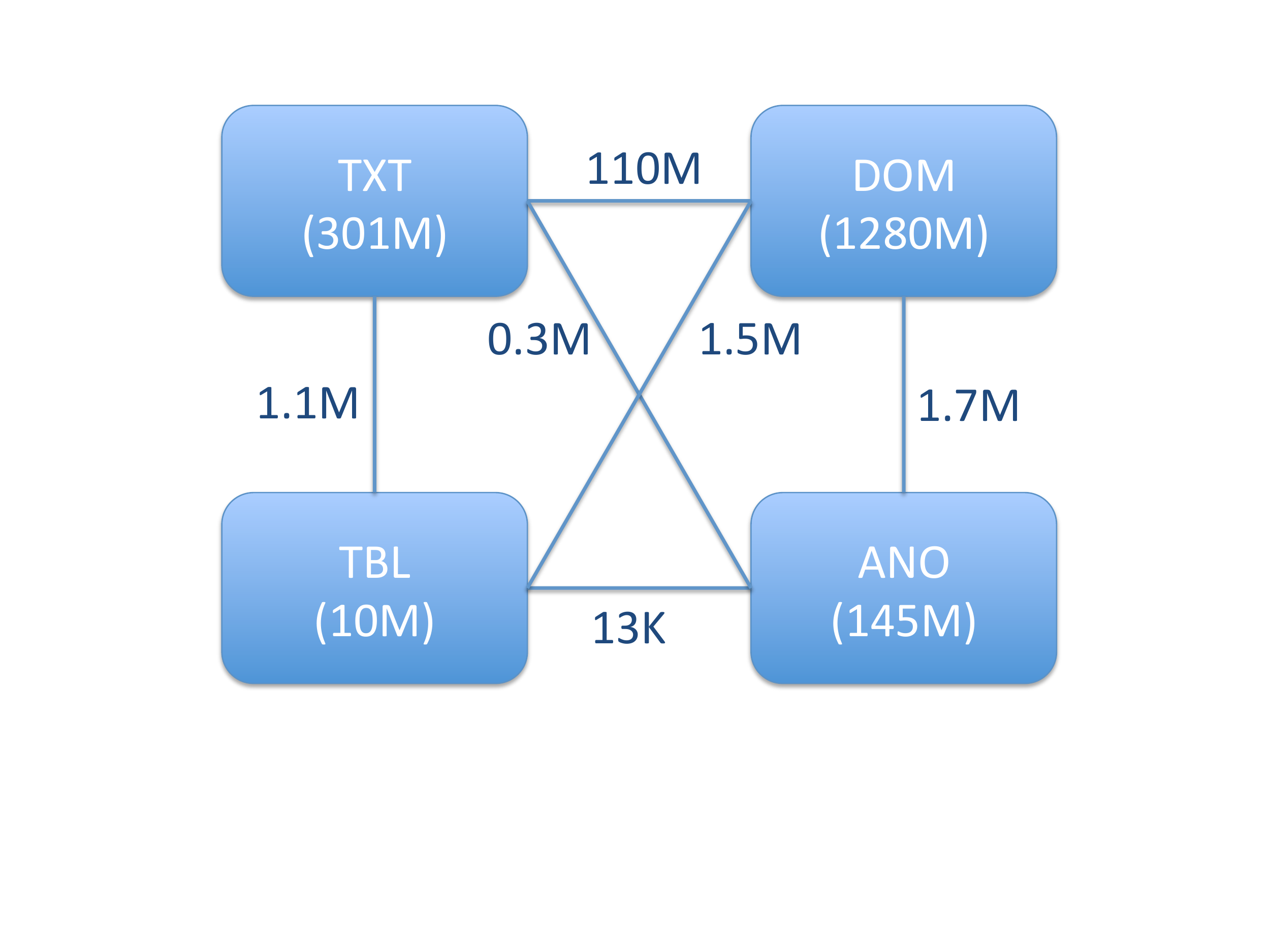}
\vspace{-.5in}
\caption{Contributions and overlaps between different types of Web contents.
\label{fig:source}}
\end{minipage}
\hfill
\begin{minipage}[th]{.32\linewidth}
\centering
\includegraphics[scale=0.3]{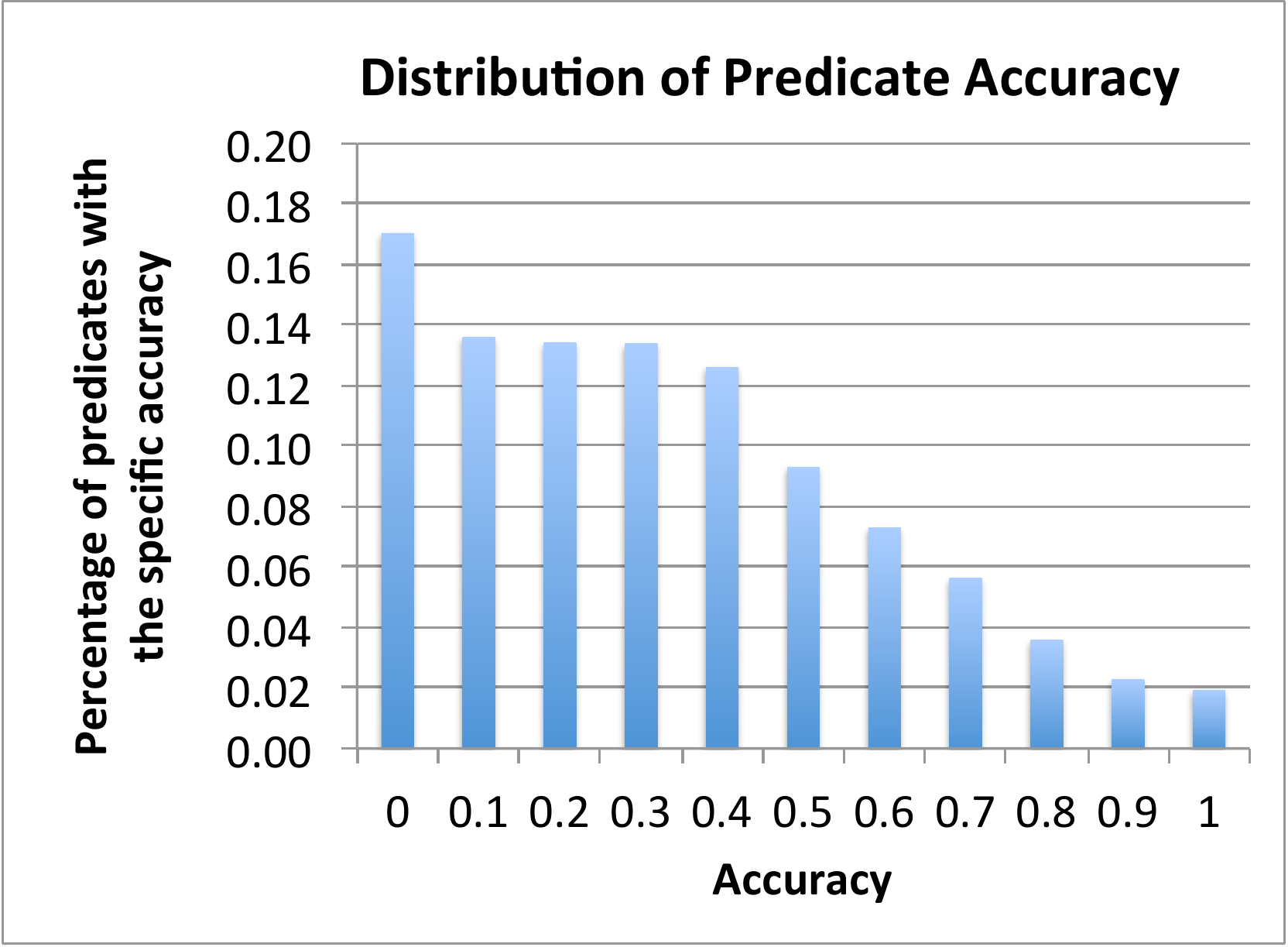}
\caption{The accuracy of extracted triples can vary a lot for different predicates.
\label{fig:accu}}
\end{minipage}
\vspace{-.2in}
\end{figure*}

Our goal is to build a high-quality Web-scale knowledge base.
Figure~\ref{fig:arch} depicts the architecture of our system;
we describe each component of this system in Section~\ref{sec:arch}.
In Section~\ref{sec:statistics} we analyze the quality of the collected knowledge,
which motivates our formal definition of the knowledge fusion problem.
While our statistical analysis is conducted on the knowledge that
we extracted using our system, we expect that similar properties will apply
to data extracted by other similar KB construction methods.

\subsection{Knowledge extraction}
\label{sec:arch}

\subsubsection{Knowledge base}
\label{sec:kb}
We follow the data format and ontology in {\em Freebase}, which contains
a large number of manually verified triples~\cite{freebase}.
We store the knowledge as ({\tt subject}, {\tt predicate}, {\tt object})
{\em triples}; an example triple is {\em (Tom Cruise, birth\_date, 7/3/1962)}.
Each {\em subject} is an {\em entity} represented by its ID in {\em Freebase},
and it belongs to one or several types;
here, we use pre-defined {\em Freebase types},
organized in a shallow 2-level hierarchy
(\eg, people/person and people/profession).
Each {\em predicate} is chosen from the set of pre-defined predicates in
{\em Freebase}. Typically a predicate is associated with a single type and can be
considered as the attribute of entities in that type.
Some predicates are {\em functional}; that is, there is a single truth for a data item
(\eg, a person has a single date of birth). Other predicates
are {\em non-functional}; for example, a person can have several children.
Each {\em object} can be an entity in {\em Freebase}, a string, or a number.
Note that in each triple the ({\tt subject}, {\tt predicate}) pair corresponds to
a ``data item'' in data fusion, and the object can be considered as
a ``value'' provided for the data item.

In this paper we focus on triples whose subjects and predicates already
exist in {\em Freebase}. Our goal is to extract new facts about these subjects
and predicates. As we show shortly, this already raises many challenges.
We note that identifying new types, new entities and new predicates
are interesting research topics and we briefly discuss these challenges
in Section~\ref{sec:improvement}.

For each extracted triple, we keep rich provenance information, including which extractor
was used to extract it, which URL the triple is extracted from, what
is the context for each extraction, and so on. Some extractors provide a confidence for
each extraction, which is kept as well.
This provenance information is much richer than
that in data fusion, which is simply the identity of the source.

\begin{table}
\centering
{\small
\caption{Overview of knowledge extracted on 10/2/2013.
The size of the data is huge
and the distribution is highly skewed in general. \label{tbl:data}}
\vspace{-.2in}
\begin{tabular}{|l|c|c|c|c|}
\hline
\#Triples & \multicolumn{4}{l|}{1.6B} \\
\#Subjects (Entities) & \multicolumn{4}{l|}{43M} \\
\#Predicates & \multicolumn{4}{l|}{4.5K} \\
\#Objects & \multicolumn{4}{l|}{102M} \\
\#Data-items (\#(sub, pred)) & \multicolumn{4}{l|}{337M} \\
\#Types & \multicolumn{4}{l|}{1.1K} \\
\hline
\hline
 & Mean & Median & Min & Max \\
\hline
\#Triples/type & 77K & 465 & 1 & 14M \\
\#Triples/entity & 38 & 8 & 1 & 2M \\
\#Triples/predicate & 370K & 323 & 1 & 191M \\
\#Triples/data-item & 4.9 & 2 & 1 & 484K \\
\#Predicates/entity & 7.8 & 5 & 1 & 618 \\
\hline
\end{tabular}
}
\vspace{-.15in}
\end{table}
\smallskip
\noindent
{\bf Statistics:} Table~\ref{tbl:data} lists basic counts for the extracted triples.
There are in total 1.6B unique triples;
together they present knowledge about 43M unique entities from 1.1K types.
The types are in various domains including geography, business, book, music, 
sports, people, biology, etc.
In total there are 4.5K unique predicates and over 300M data items.
There are many more unique objects (102M) than subjects; among them 23M are entities,
80M are raw strings (\eg, names, descriptions, addresses), and 1M are numbers.
Among the extracted triples, 83\% are not included in {\em Freebase}.

Most distributions
(\eg, \#triples per type, \#predicates per entity) are highly skewed,
which means they have a heavy head and a long tail;
indeed, we observe that the median is typically much smaller than the mean.
For example, 11 types contain millions of entities
(the top 3 types are {\sf location}, {\sf organization}, and {\sf business}),
while for 30\% types we know only up to 100 entities.
As another example, for 5 entities our knowledge consists of over 1M triples
(they are all locations: {\em USA, UK, CA, NYC, TX}, although many triples are
due to wrong extractions), while for 56\% entities we extract no more than 10 triples.

\subsubsection{Web sources}
\label{sec:sources}
We crawl a large set of Web pages and extract knowledge from four types of Web contents.
{\em Text documents (TXT)} contain texts from Web pages;
knowledge triples are typically hidden in the sentences and phrases.
For example, the sentence {\em ``Tom Cruise is an American film actor and producer''} (from {\em Wikipedia})
yields three knowledge triples:
{\em (Tom Cruise, nationality, USA)}, {\em (Tom Cruise, profession, film actor)},
and {\em (Tom Cruise, profession, film producer)}.

{\em DOM trees (DOM)} contain information organized in DOM-tree format, which can be found
either in Web pages (\eg, Web lists, Web tables), or in deep-Web sources~\cite{MKK+08}.
The structure of a DOM tree hints at the relations between the entities.
As an example, consider the following DOM tree snippet (from a Wikipedia infobox).

{\small\tt
\noindent
$<$tr$>$

$<$th$>$Born$<$/th$>$

$<$td$>$

\ \ $<$span$>$Thomas Cruise Mapother IV$<$/span$>$$<$br/$>$

\ \ $<$span$>$1962-07-03$<$/span$>$$<$br/$>$

\ \ $<$span$>$Syracuse, New York $<$/span$>$
$<$/td$>$
$<$/tr$>$
}

\vspace{-.1in}
\noindent
It contains two knowledge triples: {\em (Tom Cruise, birth\_date, 7/3/1962)} and
{\em (Tom Cruise, birth\_place, Syracuse NY)}.

{\em Web tables (TBL)} contain tabular data on the Web with essentially relational
data (as opposed to visual formatting)~\cite{CHW+08}. Typically each row in the table
represents an entity (\ie, subject) and each column represents an attribute of the
entity (\ie, predicate), and the corresponding cell represents the value of the attribute
for the entity (\ie, object). Consider the following example table.

\ \ \ \ \ \ \ \ \ \ \ \ \ \ \ \ \ \ {\small
\begin{tabular}{|c|c|c|}
\hline
Movie & Release year & Actor \\
\hline
Top Gun & 1986 & Tom Cruise \\
... & & \\
\hline
\end{tabular}
}

\smallskip
\noindent
It contains two triples: {\em (Top Gun, release\_year, 1986)} and {\em (Top Gun, actor, Tom Cruise)}.

Finally, {\em Web annotations (ANO)} contain annotations manually created
by Webmasters using ontologies defined by {\em schema.org}, {\em microformats.org} etc.
The annotations provide rich evidence for extracting knowledge.
Consider the following annotations according to {\em schema.org}:

{\small\tt
\noindent
$<$h1 itemprop="name"$>$Tom Cruise$<$/h1$>$

\noindent
$<$span itemprop="birthDate"$>$7/3/1962$<$/span$>$

\noindent
$<$span itemprop="gender"$>$Male$<$/span$>$
}

\noindent
The annotation indicates two knowledge triples: {\em (Tom Cruise, birth\_\\date, 7/3/1962)}
and {\em (Tom Cruise, gender, male)}.

\smallskip
\noindent
{\bf Statistics:} 
We crawled the Web and extracted triples from over 1B Web pages.
Figure~\ref{fig:source} shows the number of triples we extracted from each type
of Web contents. {\em DOM} is the largest contributor and contributes 80\% of the
triples, while {\em TXT} comes next and contributes 19\% of the triples.
Note however that the contribution is also bounded by the capability
of our extractors, thus only partially reflects the amount of knowledge
in a particular type of Web contents.
We may extract the same triple from different types of Web contents;
but interestingly, the overlap is rather small.
Contributions from Web sources are highly skewed: the largest Web pages
each contributes 50K triples (again, some are due to wrong extractions),
while half of the Web pages each contributes a single triple.

\subsubsection{Extractors}
\label{sec:extractors}
We employ a variety of extractors that extract knowledge triples from
the Web. There are three tasks in knowledge extraction.
The first is {\em triple identification}; that is, deciding which words or phrases describe a triple.
This is easy for certain types of Web contents such as Web tables,
but can be harder for other types of Web contents, as shown in the example
text and DOM snippet above.
The second task is {\em entity linkage}; that is, deciding which {\em Freebase} entity
a word or phrase refers to. For example, we need to decide that
{\em ``Thomas Cruise Mapother IV''} refers to the actor Tom Cruise
with {\em Freebase} ID {\em /m/07r1h}.
The third task is {\em predicate linkage}; that is, to decide which {\em Freebase} predicate
is expressed in the given piece of text. Sometimes the predicates are explicitly specified
such as in the annotation {\em itemprop=``birthDate''}.
However, more commonly, the predicates are implicit;
for example, in sentence {\em ``Tom Cruise is an American film actor and producer''},
the predicate of the triple {\em (Tom Cruise, nationality, USA)} is
implicit.
\eat{Since there can be many ways of expressing a given predicate in
natural language, a given extractor might have many different
templates or patterns that it uses internally. Consequently, the
quality of an extractor can be highly variable, as we show soon.}

Our extractors apply different techniques for different types of Web contents.
For texts, we first run standard natural language processing tools for
named entity recognition, parsing, co-reference resolution, etc.
We then apply {\em distant supervision}~\cite{Mintz09} using {\em Freebase}
triples as training data. Since there can be many ways for expressing
a given predicate in natural language, an extractor can learn many different
patterns (or templates) for each predicate. For DOM trees, we apply
distant supervision in a similar way except that we derive features
from the DOM-tree structure. For Web tables, we apply state-of-the-art
schema mapping techniques as described in~\cite{BBR11} to map table columns
to {\em Freebase} predicates. For annotations we rely on semi-automatically
defined mappings from the ontology in {\em schema.org} to that
in {\em Freebase}. For entity linkage, we use techniques
similar to~\cite{Ratinov11}.
Some extractors may extract triples from multiple types of Web contents;
for example, an extractor targeted at DOM can also extract from TBL
since Web tables are in DOM-tree format.
We omit the details for these techniques
as they are not the focus of this paper.

We note that different extractors may perform the three basic tasks
in different order; for example, some extractors first perform entity
linkage and use the results for triple identification,
while others first identify the triples and then reconcile the entities.
The three tasks may be combined; for example, some
natural language processing tools may perform triple
identification and predicate linkage at the same time.
Finally, we note that the extractors can be correlated: some may
apply the same underlying techniques and differ
only in the features selected for learning;
multiple extractors may use the same entity linkage tool.

\begin{table}
\centering
{\small
\caption{Extractors vary a lot in the number of triples they extract
and extraction quality. 
\label{tbl:extractor}}}
\vspace{-.1in}
\resizebox{\columnwidth}{!} {
\begin{tabular}{|c|c|c|c|c|c|}
\hline
 & \#Triples & \#Webpages & \#Patterns & Accu & Accu (conf $\geq .7$)\\
\hline
TXT1    & 274M & 202M & 4.8M  & 0.36 & 0.52 \\  
TXT2 & 31M  & 46M  & 3.7M  & 0.18 & 0.80    \\  
TXT3 & 8.8M & 16M  & 1.5M  & 0.25 & 0.81    \\  
TXT4 & 2.9M & 1.2M & 0.1M  & 0.78 & 0.91    \\  
\hline
DOM1 & 804M & 344M & 25.7M & 0.43 & 0.63 \\  
DOM2 & 431M & 925M & No pat. & 0.09 & 0.62 \\  
DOM3 & 45M  & N/A    & No pat.  & 0.58 & 0.93 \\  
DOM4 & 52M  & 7.8M & No pat.    & 0.26 & 0.34 \\  
DOM5 & 0.7M & 0.5M & No pat.    & 0.13 & No conf.    \\  
\hline
TBL1 & 3.1M & 0.4M & No pat.     & 0.24 & 0.24 \\  
TBL2 & 7.4M & 0.1M & No pat.     & 0.69 & No conf.    \\  
\hline
ANO & 145M & 53M  & No pat.     & 0.28 & 0.30 \\   
\hline
\end{tabular}  
}
\vspace{-.15in}
\end{table}
\smallskip
\noindent
{\bf Statistics:} 
We used 12 extractors: 4 for TXT, 5 for DOM, 2 for TBL, and 1 for ANO
(Table~\ref{tbl:extractor}); each extractor is a rather complex system, 
so we skip the details. 
Different extractors for the same type of Web contents often use broadly similar methods,
but differ in terms of their implementation details, signals used in training, and the set of
Webpages they are designed to operate on. Specifically,
TXT2-TXT4 use the same framework but run on normal Webpages, newswire, 
and {\em Wikipedia} correspondingly, whereas TXT1 has a different implementation
and runs on all Webpages. On the other hand, DOM1 and DOM2 run on all Webpages
but differ in implementation details, DOM3 and DOM4 focus on identifying entity
types, and DOM5 runs only on {\em Wikipedia}. 
Finally, TBL1 and TBL2 use different schema mapping techniques. 
We observe a high variety in the number of unique triples extracted by each extractor,
and in the number of the Webpages each extractor extracts triples from.
Some of the extractors have trained millions of patterns,
while others (\eg, extractors for Web tables) do not involve any pattern.
As we elaborate soon, we also observe a high variety in the extraction quality of the extractors.

\eat{The majority of the triples (75\%) are extracted by a single extractor,
while only 0.25\% of them are provided by more than 3 extractors.
Similarly, the majority of the Web pages (88\%) are extracted by a single extractor,
while only 0.3\% of them are extracted by more than 3 extractors.
This is mainly because we have four types of extractors, and they extract from different
types of data.
}

\subsection{Knowledge fusion}
\label{sec:statistics}

\subsubsection{Quality of extracted knowledge}
\label{sec:quality}
Evaluating the quality of the extracted triples requires a
gold standard that contains true triples and false triples.
We use {\em Freebase} for this purpose, since its triples are
manually verified and thus highly reliable.
If a triple $(s, p, o)$  occurs in {\em Freebase}, we consider it as true.
If a triple does not occur in {\em Freebase}, we could assume it is false;
this would be equivalent to making the closed-world assumption.
However, we know that this assumption is not valid, since for tail
entities, many facts are missing.
Instead, we make a slightly more refined assumption, which we call
the {\em local closed-world  assumption (LCWA)}.
Under LCWA,
if  a triple $(s,p,o)$ does not occur in {\em Freebase} but the pair $(s, p)$ does, we consider it as false;
but if $(s,p)$ does not occur in {\em Freebase}, we abstain from labeling
the triple, and exclude it from the gold standard.
In other words, we assume once {\em Freebase} has knowledge about a particular data item,
it has complete knowledge, so it is locally complete.

This assumption is valid for functional predicates,
but not for non-functional ones. This is because
{\em Freebase} may only contain a subset of true triples for a data item,
and hence we could wrongly label the extra true triples we have extracted as false.
For example, the set of actors in a movie is often
incomplete in {\em Freebase}. Nevertheless, in practice
we observe that most predicates have only 1 or 2 true values, so the LCWA assumption
works well. Note also that \cite{GTH+13} 
has shown the effectiveness of LCWA in data extraction.

Among the 1.6B triples, 650M (40\%) have gold standard labels,
of which 200M are labeled as correct (\ie, included in {\em Freebase});
in other words, the accuracy of the extracted
triples is estimated to be about 30\% (assuming the accuracy of
triples not in the gold standard is similar to those in the
gold standard).
Note that this includes triples of different extraction confidence;
as we show shortly, oftentimes triples with a lower confidence have lower accuracy,
but the correlation can be different for different extractors.
Figure~\ref{fig:accu} shows that the accuracy varies from predicate to predicate:
44\% of the predicates have very low accuracy (below 0.3),
while 13\% of the predicates have fairly high accuracy (above 0.7).

Some of the erroneous triples
are due to wrong information provided by Web sources whereas others are due to mistakes
in extractions. Unfortunately, it is hard to distinguish them,
unless we can automatically decide whether a source indeed claims a particular
knowledge triple. A random sampling of 25 false triples shows that
44\% have triple-identification errors,
such as taking part of the album name as the artist for the album;
 44\% have entity-linkage errors, such as wrongly reconciling
the Broadway show {\em Les Miserables} to the novel of the same name;
20\% have predicate-linkage errors, 
such as mistaking the book author as the book editor; 
and only 4\% are indeed provided by the Web sources
(some triples are due to multiple errors).
Thus, extractions are responsible for the majority of the errors.

\begin{figure*}
\vspace{-.1in}
\begin{minipage}[th]{.32\linewidth}
\centering
\includegraphics[scale=0.3]{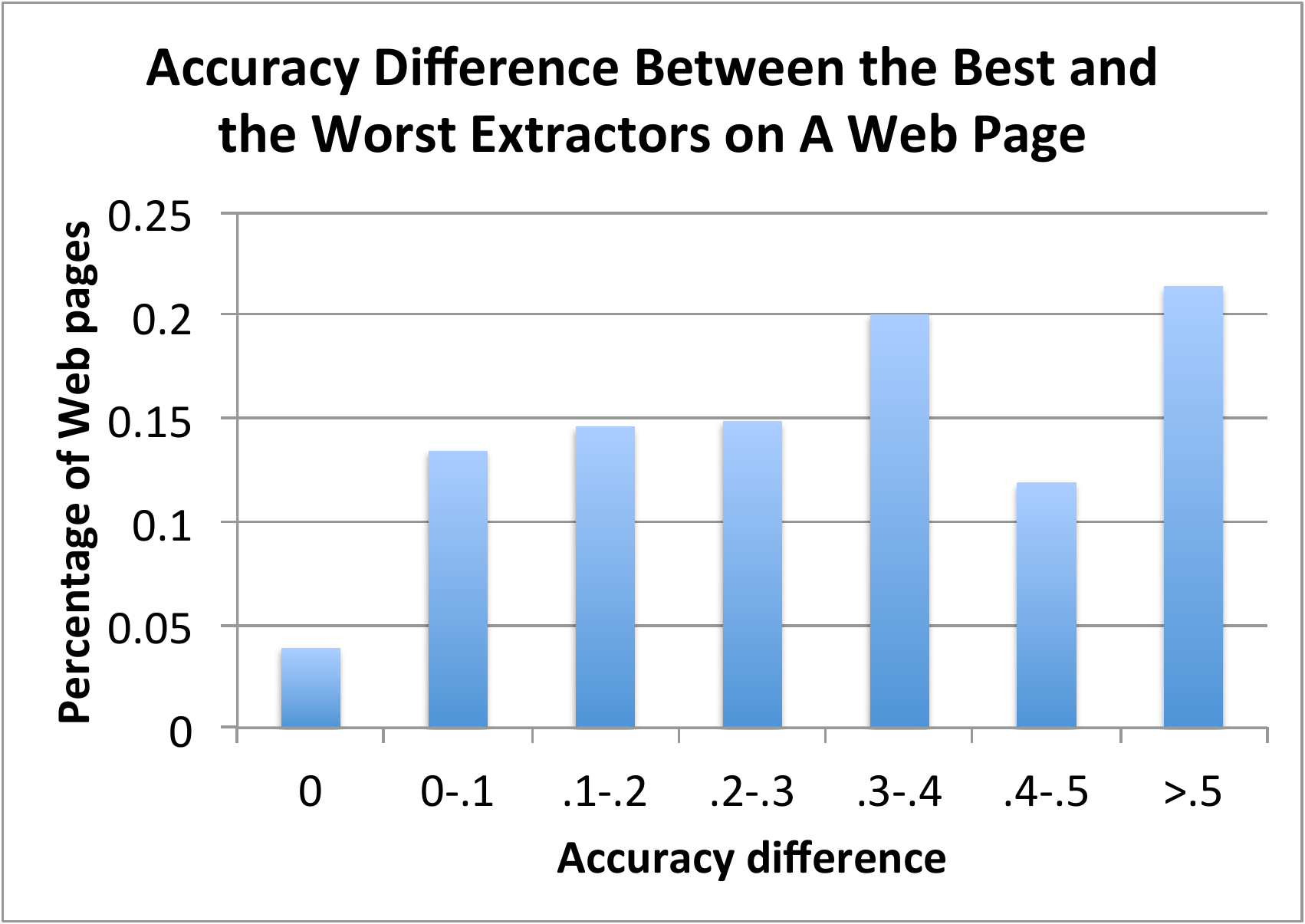}
\caption{
For a lot of Web sources extraction quality varies a lot for different extractors.
\label{fig:extractor}}
\end{minipage}
\hfill
\begin{minipage}[th]{.32\linewidth}
\centering
\includegraphics[scale=0.3]{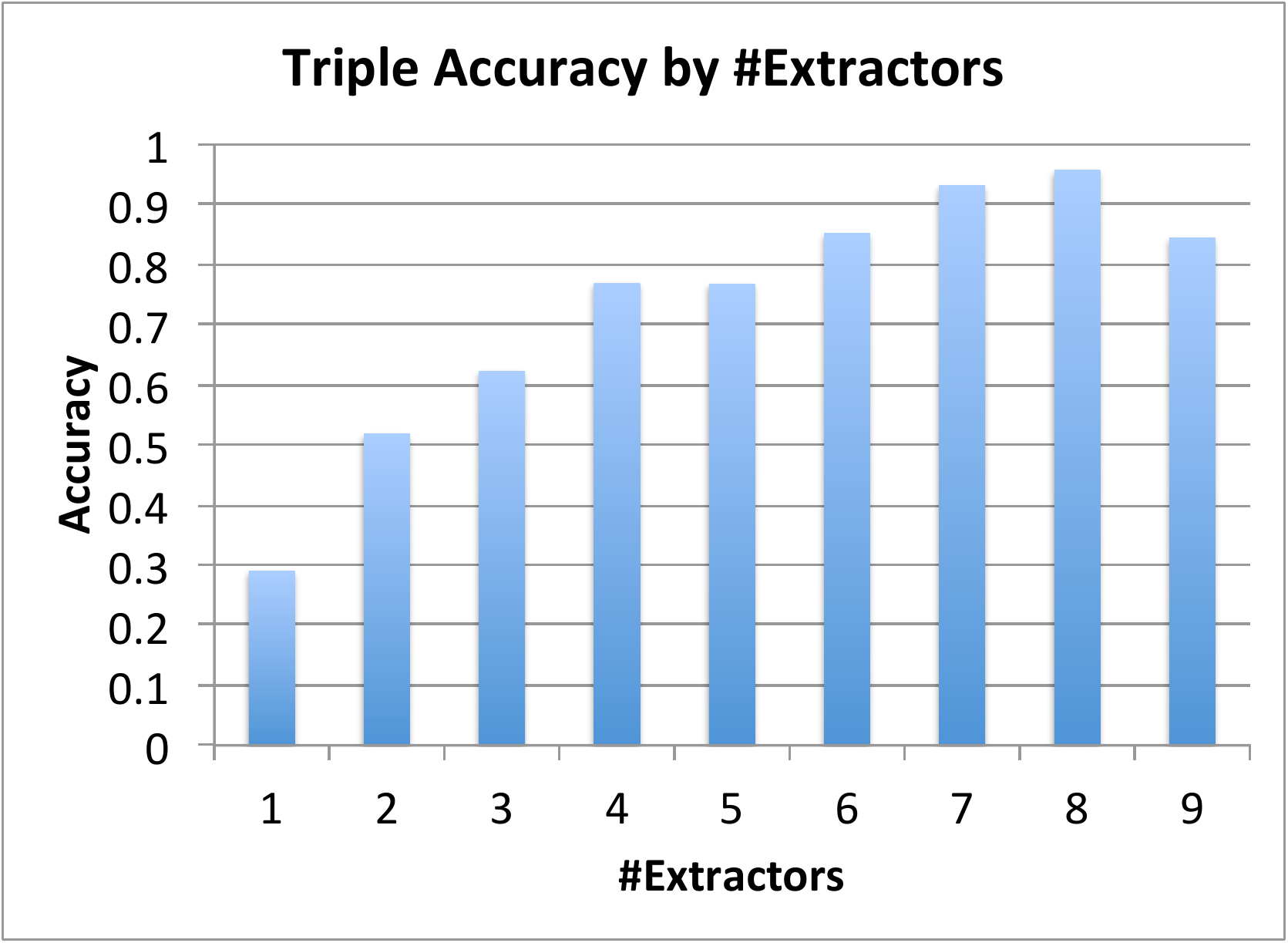}
\caption{Accuracy of triples
increases with \#Extractors, but there are drops sometimes.
\label{fig:accuByExt}}
\end{minipage}
\hfill
\begin{minipage}[th]{.32\linewidth}
\centering
\includegraphics[scale=0.3]{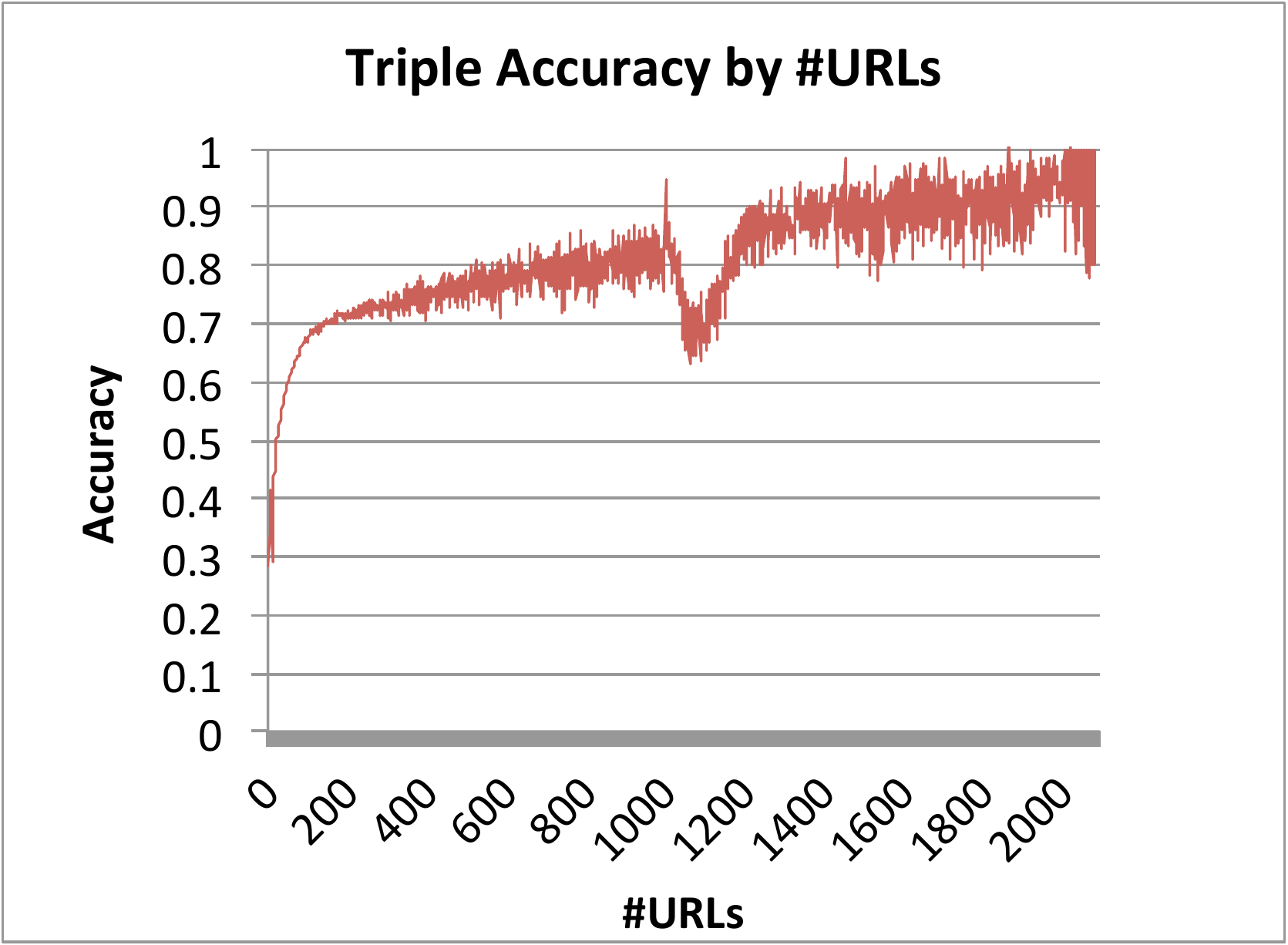}
\caption{Accuracy of triples
increases with \#Sources, but there are drops sometimes.
\label{fig:accuBySrc}}
\end{minipage}
\vspace{-.15in}
\end{figure*}

As shown in Table~\ref{tbl:extractor}, the accuracy of the extractors
ranges from 0.09 to 0.78 (again, some errors are due to
wrong information provided by the Web sources).
Within each extractor, we observe high variance of extraction
quality among different patterns, and on different Web pages;
in most cases the accuracy ranges from nearly 0 to nearly 1
under the same extractor.
For extractors that attach a confidence to each extraction,
typically we observe a higher accuracy for triples with a high
confidence (\ie, normalized confidence above .7); however, the variance differs
for different extractors. 

It is hard to evaluate the trustworthiness of a Web source,
since many errors from the Web source are actually introduced by the extractors.
We measure each Web source by the quality of triples extracted by a particular
extractor; it turns out we often obtain very different quality measures when we
use different extractors. (Here we consider an extractor
for a Web source only if it extracts at least 5 triples from that source.)
As shown in Figure~\ref{fig:extractor},
for a Web page the difference between the accuracy of the best extractor 
and that of the worst one is 0.32 on average, and above 0.5 
for 21\% of the Web pages.

Finally, as shown in Figures~\ref{fig:accuByExt}-\ref{fig:accuBySrc},
the more Web sources from which we extract a triple,
or the more extractors that extract a triple,
the more likely the triple is true. But there can be exceptions:
in Figure~\ref{fig:accuByExt} there is a big drop when the number 
of extractors increases from 8 to 9, mainly because of correlations
between the extractors;
in Figure~\ref{fig:accuBySrc} the curve fluctuates and there is
a big drop when the number of sources is in $[1K, 1.1K)$,
mainly because of common errors by the same extractor on many different sources.
On the other hand, triples that lack significant support can still have a good chance to be true:
the majority of the triples (75\%) are extracted by a single extractor
and their accuracy is nearly 30\%;
half of the triples (51\%) are extracted from a single Web page
and their accuracy is also nearly 30\%.

\subsubsection{Knowledge fusion}
\label{sec:defn}
The many errors in the extracted candidate triples call for a strategy
that can automatically decide the correctness for each
triple; that is, whether the triple is consistent with the real world.
Instead of making binary decisions and discarding the triples that are
determined to be false, we wish to output for each triple a truthfulness probability
between 0 and 1. We can use such probabilistic knowledge in three ways.
For the triples with very high probabilities, we can trust them and use them
directly in applications. For the triples with very low probabilities,
we can feed them to extraction systems as negative training examples.
The triples with medium probabilities can be
used as input to  active learning strategies.
Motivated by this, we now formally define the knowledge fusion problem.

\begin{definition}[Knowledge fusion]
Given a set of extracted knowledge triples, each associated with provenance information
such as the extractor and the Web source, {\em knowledge fusion}
computes for each unique triple the probability that it is true. \rbox

\eat{
Let $\bf T$ be a set of extracted triples, each in the form of $(t, e, s, c)$,
where $t$ is a triple containing (subject, predicate, object), $e$ is the
extractor that extracts the triple, $s$ is the Web source from which the triple
is extracted, and $c$ indicates the confidence of extraction ($c$ can be null).
Let $\bf G$ be a set of ground truths, each in the form of $(t, b)$, where
$t$ is a triple and $b$ is a label indicating whether $b$ is true.
Suppose $\bf G$ may contain extra triples that are not in $\bf T$.
Knowledge fusion decides for each unique triple $t$ from $\bf T$,
the probability that $t$ is true. \rbox}
\end{definition}

Compared with data fusion, knowledge fusion raises three challenges.
First, recall that data fusion takes as input a two-dimensional data matrix;
in contrast, the input of knowledge fusion is three-dimensional (Figure~\ref{fig:input}(b)).
The additional dimension represents extractors, so each cell in the matrix represents
what the corresponding extractor extracts from the corresponding Web source
on the corresponding data item. In other words, a triple is not necessarily
what is provided by a source, but what is extracted by an extractor from the source.
Errors can creep in at every stage in this process, not only from the Web sources, but also
from triple identification, entity linkage, and predicate linkage by the extractors.

Second, recall that data fusion makes a binary decision on which value
is true for a data item; in contrast, the output of knowledge fusion is
a truthfulness probability for each triple.
We wish the predicted probability to truly reflect the likelihood
that the triple is true. A basic requirement is {\em monotonicity}:
a triple with a higher predicted probability should be more likely to be true
than a triple with a lower predicted probability. A more demanding requirement is
{\em calibration}: among all triples predicted with a certain probability $p$,
ideally a $p$ fraction of them are indeed true.

Third, the scale of knowledge is typically huge.
Currently the largest data sets used in data fusion experiments contain
up to 170K data sources~\cite{PR13}, up to 400K data items,
and up to 18M provided triples~\cite{LDL+12}. Knowledge fusion often needs to
handle data that is orders of magnitude larger in every aspect.
The overall size of our data is 1.1TB;
there are 12 extractors, 1B+ Web sources, 375M data items,
1.6B unique triples, and 6.4B extracted triples.

\eat{
First, a triple is not necessarily what is provided by a source, but what
is extracted by an extractor from a source; we show shortly the challenge
of having extractors as a middle layer between the raw data and the knowledge.
Second, as a result, we have richer provenance information, including not
only the data provider, but also the extractor, sometimes extraction
contexts and confidence. Third, our goal is to compute a calibrated probability
for truthfulness, rather than to make a binary decision.
}

\section{Applying Data Fusion Methods to Knowledge Fusion}
\label{sec:performance}
We apply existing data fusion (DF)
methods to solve the knowledge fusion (KF) problem.
In Section~\ref{sec:adapt} we describe our adaptations.
In Section~\ref{sec:basic}, we evaluate the quality of these methods,
and note several problems.
In Section~\ref{sec:refine}, we suggest some simple improvements
that improve the quality considerably.
Finally, in Section~\ref{sec:err}, we analyze the remaining errors
made by our system.


\subsection{Adapting data fusion techniques}
To select from existing DF methods, we adopted three criteria.
First, since the goal of KF
is to compute for each triple a truthfulness probability,
we select DF methods that can easily derive a meaningful probability.
This rules out methods such as
\cite{YHY07}, which
computes a probability close to 1 for every triple,
as well as Web-link based methods and IR-based methods,
which compute correctness scores that do not have a probabilistic interpretation.

Second, since the scale of our data is three orders of
magnitude larger than that used in traditional DF, we select methods that can be fairly
easily scaled up under a MapReduce based framework~\cite{mapreduce}.
Scaling up some DF methods is non-trivial.
For example, relation-based methods reason about every pair
or even every subset of data sources; this is prohibitively expensive for 
the 1B+ Web sources in our data set even with parallelization.
Similarly, we have not found an easy way to scale up graphical-model methods.

Third, we focus on methods for which latest studies have shown great promise;
for example, \cite{LDL+12} has shown the advantages of Bayesian methods
over Web-link based and IR-based methods.

This selection process yields three methods: {\sc Vote}, {\sc Accu},
and {\sc PopAccu}. We describe them briefly,
and then describe how we adapt them to the KF problem setting.

\smallskip
\noindent
{\sc Vote}: For each data item, {\sc Vote} counts the sources for each value
and trusts the value with the largest number of sources.
{\sc Vote} serves as a baseline in our experiment.

\eat{enhances of each associated unique triple,
and assign a proportional probability. Specifically, if a data item $D=(s, p)$ has $n$ provenances
and a triple $t=(s, p, o)$ has $m$ provenances, the probability of $t$ is $p(t)={m over n}$.
}

\smallskip
\noindent
{\sc Accu}: {\sc Accu} applies Bayesian analysis proposed in~\cite{DBS09a}.
For each source $S$ that provides a set of values ${\bf V}_S$, the accuracy of
$S$ is computed as the average probability for values in ${\bf V}_S$.
For each data item $D$ and the set of values ${\bf V}_D$ provided for $D$, the probability of
a value $v \in {\bf V}_D$ is computed as its {\em a posterior} probability conditioned
on the observed data using Bayesian analysis.
{\sc Accu} assumes that (1) for each $D$ there is a single true value,
(2) there are $N$ uniformly distributed false values, and
(3) the sources are independent of each other.
We start with a default accuracy $A$ for each source 
and iteratively compute value probability and source accuracy.
By default we set $N=100$ and $A=0.8$.
For lack of space, we omit the equations and refer our readers to~\cite{DBS09a} for details.

\smallskip
\noindent
{\sc PopAccu}: {\sc PopAccu}~\cite{DSS13} extends {\sc Accu} by removing the assumption
that wrong values are uniformly distributed; instead, it computes the distribution from
real data and plugs it in to the Bayesian analysis. It is proved in~\cite{DSS13} that
{\sc PopAccu} is monotonic in the sense that adding a data source would not decrease
the quality of data fusion under the assumption that all sources and data items
are independent; it is also empirically
shown in~\cite{DSS13, LDL+12} that {\sc PopAccu} is more robust than {\sc Accu}
in case there exists copying between the sources, because copied false values may be
considered as {\em popular} false values.

\begin{figure}
\centering
\includegraphics[scale=0.35]{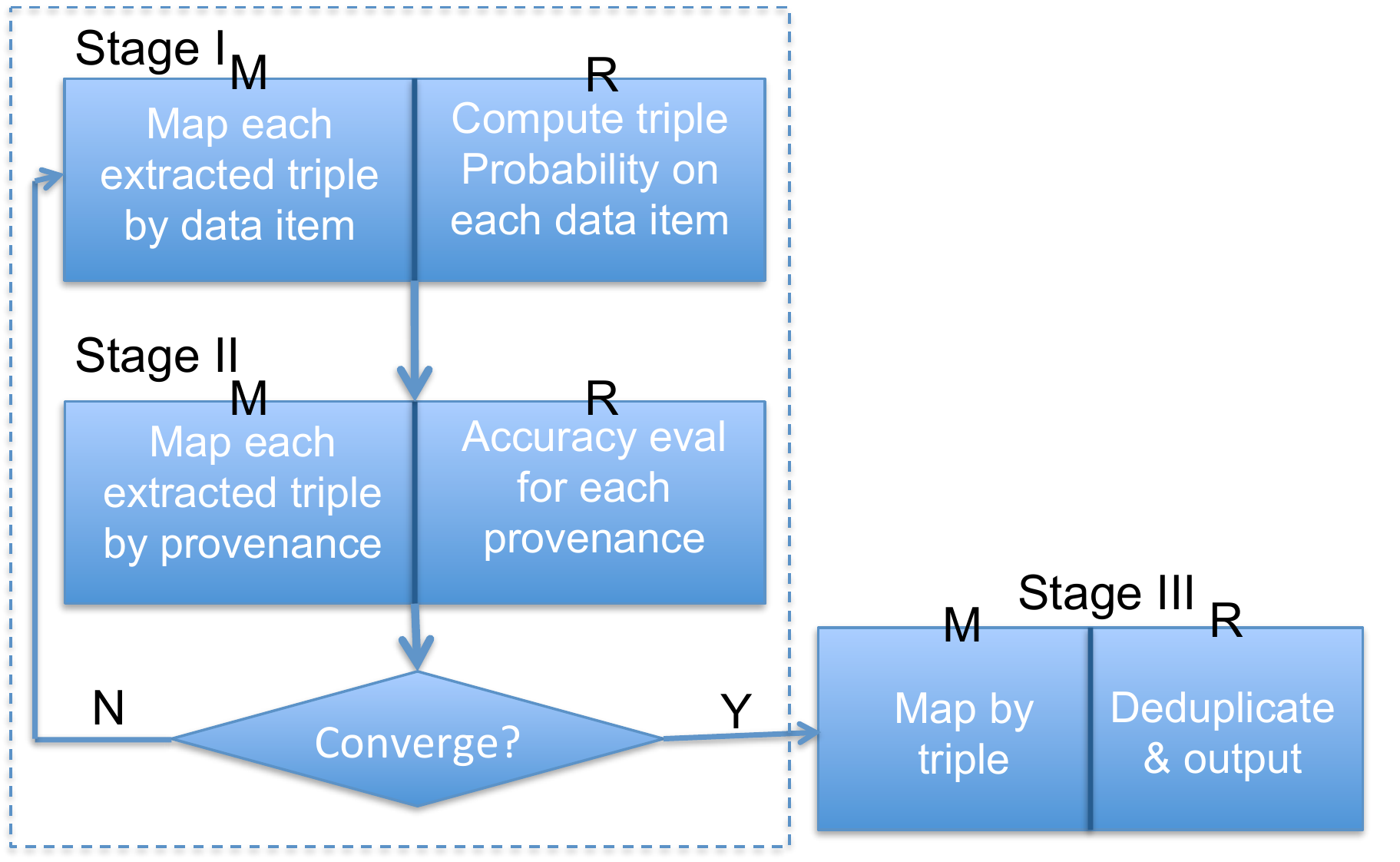}
\vspace{-.1in}
\caption{MapReduce implementation of {\sc Accu} and {\sc PopAccu}.
\label{fig:mapreduce}}
\vspace{-.2in}
\end{figure}
\label{sec:adapt}
\smallskip
\noindent
{\bf Adaptations:}
We adapt the data fusion methods in three ways to solve the knowledge fusion problem.

First, recall that the input of DF methods is a two-dimensional data matrix with
values provided by each source on each data item, whereas the input of KF methods
is three-dimensional, containing the values extracted by each extractor from each
source on each data item. We reduce the dimension of the KF input by considering
each (Extractor, URL) pair as a data source, which we call a {\em provenance}.
Having a large number of provenances indicates either the triple is supported by
many Web sources, or the triple is extracted by many different extractors;
both presumably would increase our confidence in the correctness of the triple.
We describe shortly how we may vary the granularity of the provenance.

Second, recall that the output of the DF methods consists of binary decisions on each
provided value, whereas the output of the KF methods consists of truthfulness
probabilities for each triple. For {\sc Accu} and {\sc PopAccu}, we simply
take the probability computed by the Bayesian analysis \cite{DSS13}. For {\sc Vote}, we
assign a probability as follows:
if a data item $D=(s, p)$ has $n$ provenances in total
and a triple $T=(s, p, o)$ has $m$ provenances, the probability of $T$ is
$p(T)={m \over n}$. Note that all three methods assume
single-truth: the probabilities of different triples associated with the same
data item sum up to 1.
This assumption is theoretically invalid for non-functional predicates,
but in practice it performs surprisingly well, as we discuss shortly.

\eat{
{\small
\begin{algorithm}[t]
\caption{{\sc PopAccu}($\bf ET$)\label{algo:mapreduce}}
\Input{$\bf ET$: collection of extracted triples, each in the form of (triple, provenance)}
\Output{$\bf pT$: Triples with predicted truthfulness probability}

\Repeat {there is no big accuracy change}{
\tcp{Stage I.}
{\sc Map}: Map each extracted triple by the data item; \\
{\sc Reduce}: compute probability for each triple of the same data item; \\

\tcp{Stage II.}
{\sc Map}: Map each extracted triple by the provenance; \\
{\sc Reduce}: compute provenance accuracy;
}

\tcp{Stage III.}
{\sc Map}: Map each extracted triple by the triple; \\
{\sc Reduce}: Remove duplicates from different provenances; \\
\Return {unique triples with probabilities};

\end{algorithm}
}}

Third, we scale up the three methods using a MapReduce-based framework:
Figure~\ref{fig:mapreduce} shows the architecture for knowledge fusion.
There are three stages; each stage is a MapReduce process and so is performed
in a parallel fashion. In the first stage, the {\em Map} step takes as input
the {\em extracted triples}, each being a (triple, provenance) pair,
and partitions them by the associated data item; the {\em Reduce} step
applies Bayesian inference on
all triples provided for the same data item and computes the probability
for each of them. 
In the second stage, the {\em Map} step partitions each extracted
triple, where the triple is already assigned a probability, by its provenance;
the {\em Reduce} step computes the accuracy of each provenance
from its triples. We iterate the first two stages
until convergence. The third stage outputs the results: the {\em Map} step partitions
extracted triples by the triple and the {\em Reduce} step removes duplicates from different provenances.
Note that {\sc Vote} does not need the iterations and has only Stage I and Stage III.

We use two additional strategies to speed up execution.
First, recall that we observe a big skew in the data: the number of extracted triples
for each data item ranges from 1 to 2.7M, and the number of triples a
provenance contributes ranges from 1 to 50K.
To speed up computation at each reducer, whenever applicable,
we sample $L$ triples (by default we set $L=1M$) each time instead of using all triples
for Bayesian analysis or source accuracy evaluation.
Second, there might be many rounds before convergence and even a single round can take
a long time. We force termination after $R$ rounds (by default we set $R=5$).
We examine the effect of these choices on performance shortly.

\begin{figure*}
\vspace{-.1in}
\begin{minipage}[th]{.32\linewidth}
\centering
\includegraphics[scale=0.3]{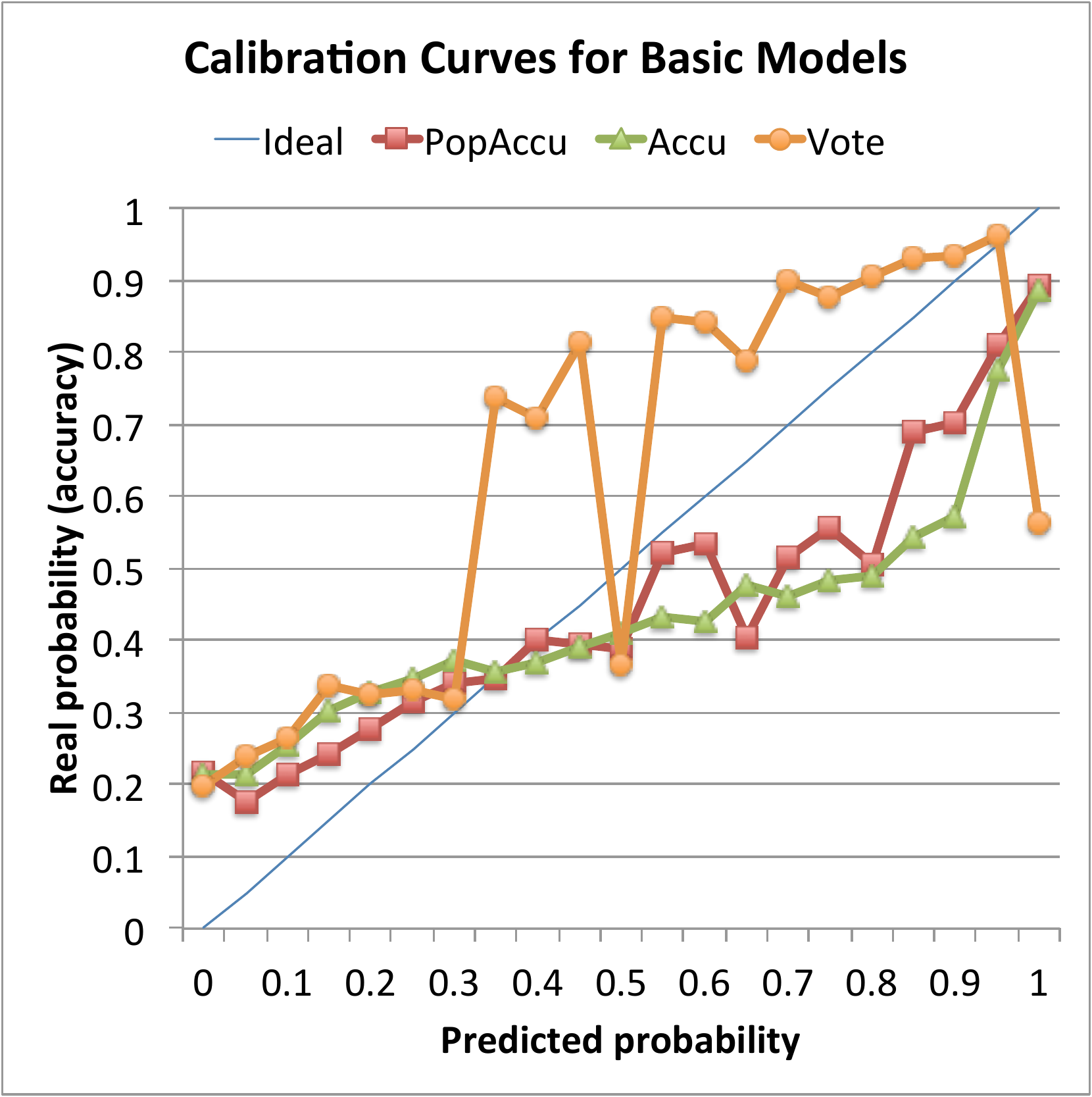}

\ \

\resizebox{\columnwidth}{!} {
\begin{tabular}{|c|c|c|c|}
\hline
 & Dev. & WDev. & AUC-PR \\
\hline
{\sc Vote} & .047 & .061 & .489 \\
{\sc Accu} & .033 & .042 & .524 \\
{\sc PopAccu} & {\bf .020} & {\bf .037} & .499 \\
\hline
{\sc PopAccu} (Only ext) & .049 & .052 & {\bf .589} \\
{\sc PopAccu} (Only src) & .024 & .039 & .528 \\
\hline
\end{tabular}
}

\caption{Among the three basic fusion methods, {\sc PopAccu} has the
best probability calibration. The best measures are in bold.
\label{fig:basic}}

\end{minipage}
\hfill
\begin{minipage}[th]{.32\linewidth}
\centering
\includegraphics[scale=0.3]{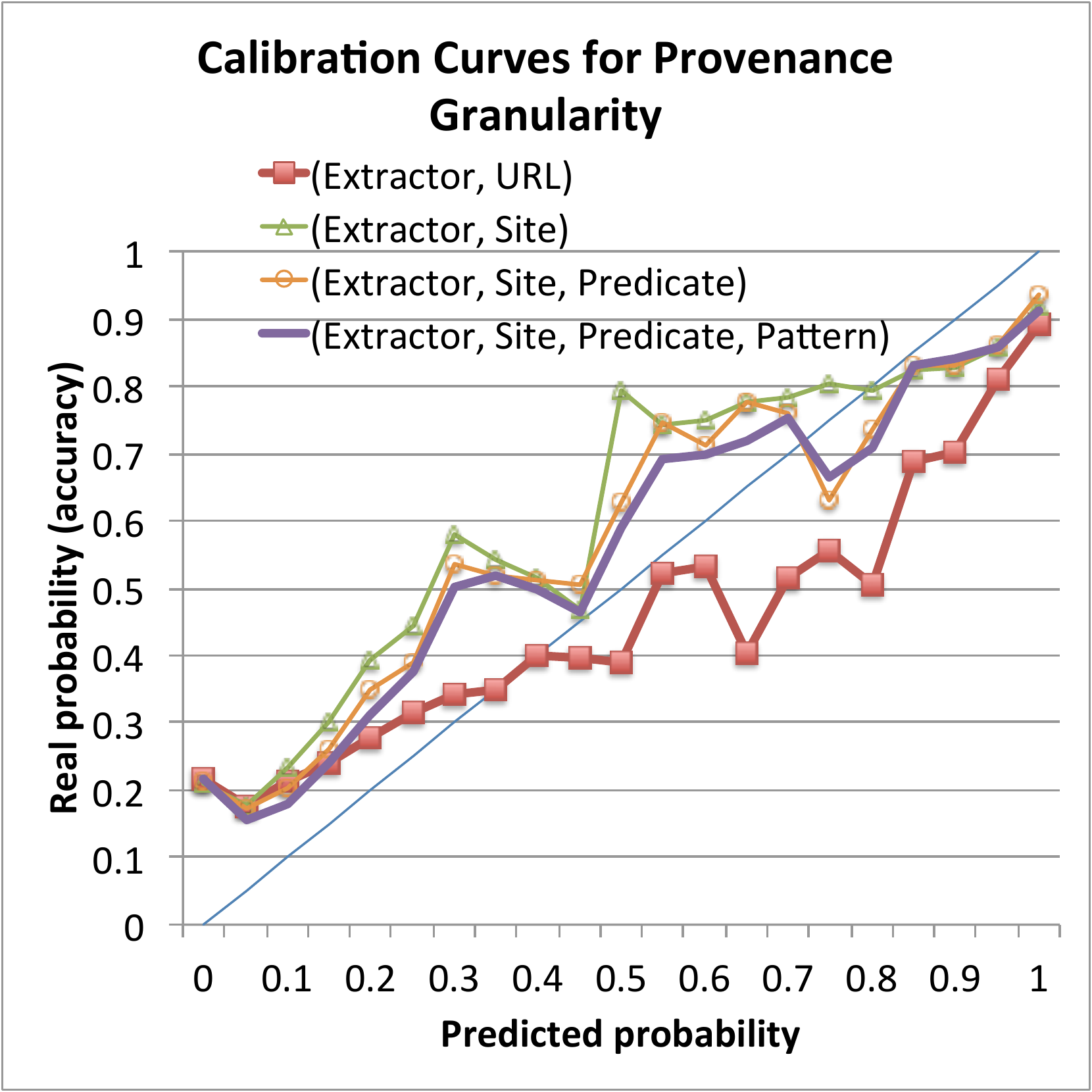}

\ \

\resizebox{\columnwidth}{!} {
\begin{tabular}{|c|c|c|c|}
\hline
 & Dev. & WDev. & AUC-PR \\
\hline
(Extractor, URL) & .020 & .037 & .499 \\
(Extractor, Site) & .023 & .042 & .514 \\
(Ext, Site, Pred) & .017 & .033 & {\bf .525} \\
(Ext, Site, Pred, Pattern) & {\bf .012} & {\bf .032} & .522 \\
\hline
\end{tabular}}

\vspace{.15in}
{\small
\caption{The granularity of (Extractor, Site, Predicate, Pattern) for
provenances obtains the best probability calibration.
\label{fig:granularity}}
}
\end{minipage}
\hfill
\begin{minipage}[th]{.32\linewidth}
\centering
\includegraphics[scale=0.3]{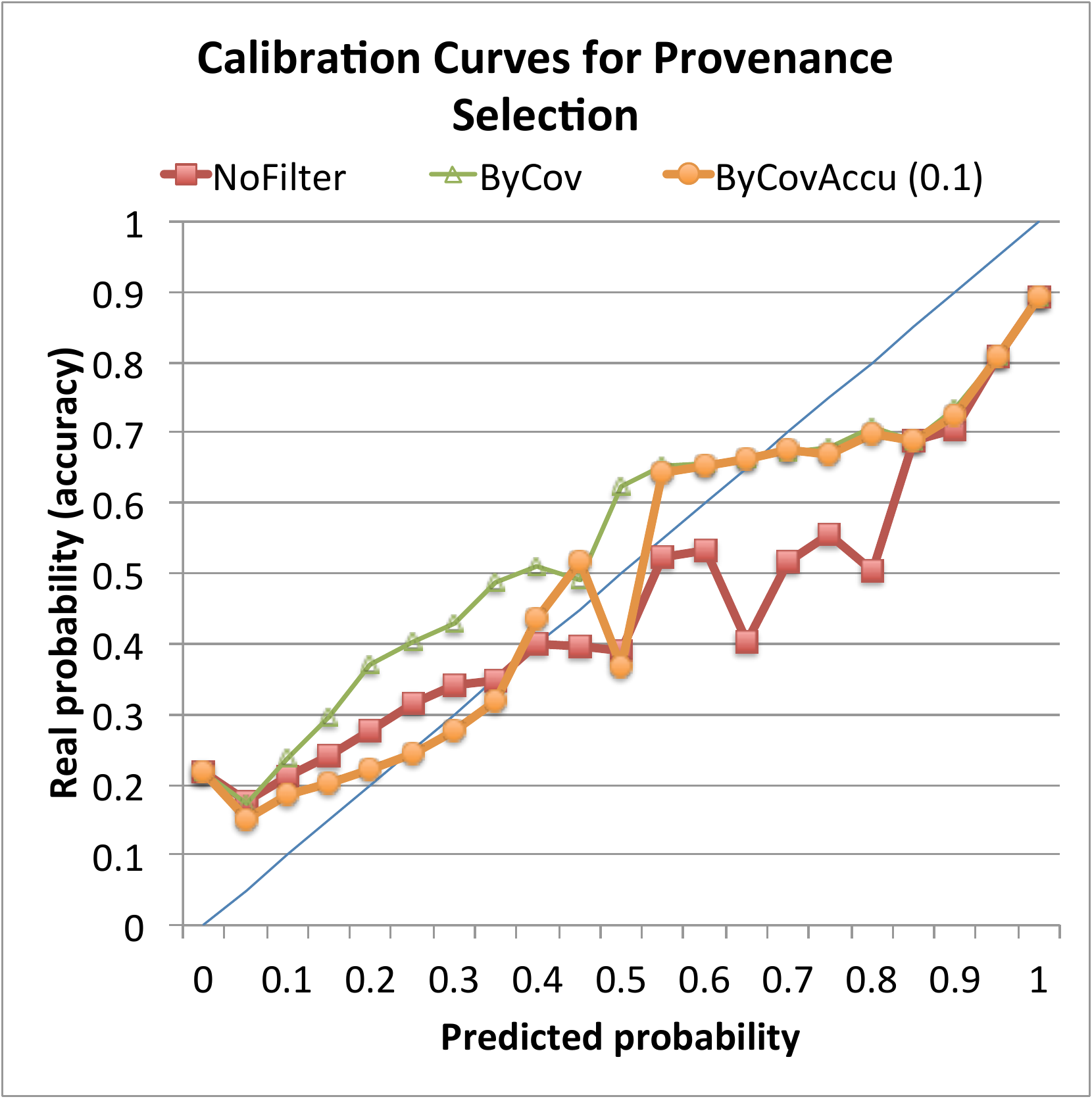}

\ \

\resizebox{\columnwidth}{!} {
\begin{tabular}{|c|c|c|c|}
\hline
 & Dev. & WDev. & AUC-PR \\
\hline
{\sc NoFiltering} & .020 & .037 & .499 \\
{\sc ByCov} & .016 & .038 & .511 \\
{\sc ByCovAccu} ($\theta=.1$) & {\bf .010} & {\bf .035} & .495 \\
{\sc ByCovAccu} ($\theta=.3$) & .017 & .038 & .516 \\
{\sc ByCovAccu} ($\theta=.5$) & .018 & .038 & {\bf .520} \\
{\sc ByCovAccu} ($\theta=.7$) & .024 & .039 & .518 \\
{\sc ByCovAccu} ($\theta=.9$) & .042 & .038 & .510 \\
\hline
\end{tabular}}
\vspace{-.05in}
{\small
\caption{Filtering sources by coverage and accuracy improves probability calibration.
\label{fig:filter}}
}
\end{minipage}
\vspace{-.2in}
\end{figure*}

\subsection{Experimental evaluation}
\label{sec:basic}

\eat{We implemented the various methods in Flume Java~\cite{CRP+10} and
used up to 1000 machines with a 4G RAM at the Google Borg job scheduling
and monitoring system~\cite{borg}.
}


\noindent
{\bf Metrics:} We use two evaluation metrics: the area
under the precision-recall curve, and a measure of how well-calibrated
our probability estimates are. We describe both of these in more
detail next.

\smallskip
\noindent
{\em PR curve:} We order the triples in decreasing order of the predicted probability.
As we gradually add new triples, we plot the precision versus the recall of
the considered triples. A good method should have a high
precision as we fix recall, and have a high recall as we fix precision.
We summarize using the area under the curve, which we call AUC-PR.

Note that PR-curves
show whether the predicted probabilities are {\em monotonic} but cannot show
whether they are {\em calibrated}: if two systems
order the triples in the same way, even if one predicts
probabilities in $[0, 0.1]$ (implying that all triples are likely to be false)
and the other predicts probabilities in $[0.9, 1]$
(implying that all triples are likely to be true),
we would observe the same PR-curve.
For this reason, we also consider an additional metric, described next.

\smallskip
\noindent
{\em Calibration curve}: We plot the predicted probability versus the real probability.
    To compute the real probability, we divide the triples into $l+1$ buckets: the $i$-th
    ($0 \leq i < l$) bucket contains triples with predicted probability in
    $[{i \over l}, {i+1 \over l})$, and the $(l+1)$-th bucket contains triples
    with probability 1.
    We use $l=20$ when we report our results. We compute the real probability
    for each bucket as the percentage of true triples in the bucket compared with our gold
    standard. Ideally the predicted probability
    should match the real probability so the ideal curve is a diagonal line from $(0,0)$ to
    $(1,1)$. Note that such curves have also been plotted in~\cite{Niu12, Wick13akbc}.

    We summarize the calibration using two measures.
    The {\em deviation} computes the average square loss between
    predicted probabilities and real probabilities for each bucket.
    The {\em weighted deviation} is the same except that it weighs each bucket by the
    number of triples in the bucket, thus essentially computes the average
    square loss of each predicted probability.

\medskip
\noindent
{\bf Results:} We evaluate the three 
methods on our extracted knowledge.
Figure~\ref{fig:basic} plots the calibration curves
and summarizes the deviation, weighted deviation, and AUC-PR
for each method. Among them, {\sc PopAccu} has the lowest weighted deviation,
then comes {\sc Accu}, and both are much better than {\sc Vote}.
In terms of PR-curves, {\sc Accu} has the highest AUC-PR, while
{\sc Vote} has the lowest.

{\sc Vote} has two inherent problems. First, except a few points, it 
under-estimates (\ie, the predicted probability is lower than the real probability)
most of the time for true triples. To understand why, consider an example data item for which
there are four unique triples, one extracted from 7 provenances while each of the others
extracted from 1 provenance. {\sc Vote} would compute a probability of
${7 \over 10} = 0.7$ for the first
triple, although intuitively the likelihood for this triple to be true is much higher.
Second, {\sc Vote} falls short when there is only 1 or 2 provenances 
for a data item, where it would compute a probability of 1 or 0.5.
However, intuitively this is often the case when the extractions go wrongly
and as shown in Figure~\ref{fig:basic}, the real accuracy for triples
with predicted probability of 1 or 0.5 is much lower
(0.56 and 0.37 respectively).

The calibration curves for {\sc Accu} and {\sc PopAccu} are better,
but they reveal two major problems.
First, we observe that they over-estimate for triples with
a high predicted probability ($\geq 0.4$), so there are a lot of false positives;
they also underestimate for triples with a low predicted
probability, so there are a lot of false negatives.

Second, we observe that the calibration curve for {\sc PopAccu} is not smooth:
there are valleys when the predicted probability is 0.8 or 0.5.
Such probabilities are mostly predicted when small provenances exist,
which is common in practice. As an example, consider the case when 
a data item is extracted from a single provenance
(so there is no conflict) and that provenance contributes a single triple.
Recall that initially we assign a default accuracy of 0.8 for each provenance;
over the iterations that single triple would carry this default accuracy
as its probability, reinforcing the default accuracy for that provenance.
Note that {\sc Accu} can also be biased in such situations, but we do not observe
obvious valleys in its curve. This is because {\sc Accu} assumes that
there are $N$ false triples in the domain for each data item,
so the computed probabilities would not stick to a particular value.

Finally, we also show in Figure~\ref{fig:basic} the results of applying
{\sc PopAccu} when we consider only extractor patterns or only URLs instead
of (Extractor, URL) pairs. Considering only extractor patterns under-estimates
most of the time as it may compute a low probability for triples
extracted by one pattern but from many sources (using extractors
instead of patterns exacerbates the problem). Considering
only sources over-estimates more when the predicted
probability is high, and under-estimates more when the predicted
probability is low, as it ignores the supports from different extractors.

\begin{figure*}
\vspace{-.1in}
\begin{minipage}[th]{.32\linewidth}
\centering
\includegraphics[scale=0.3]{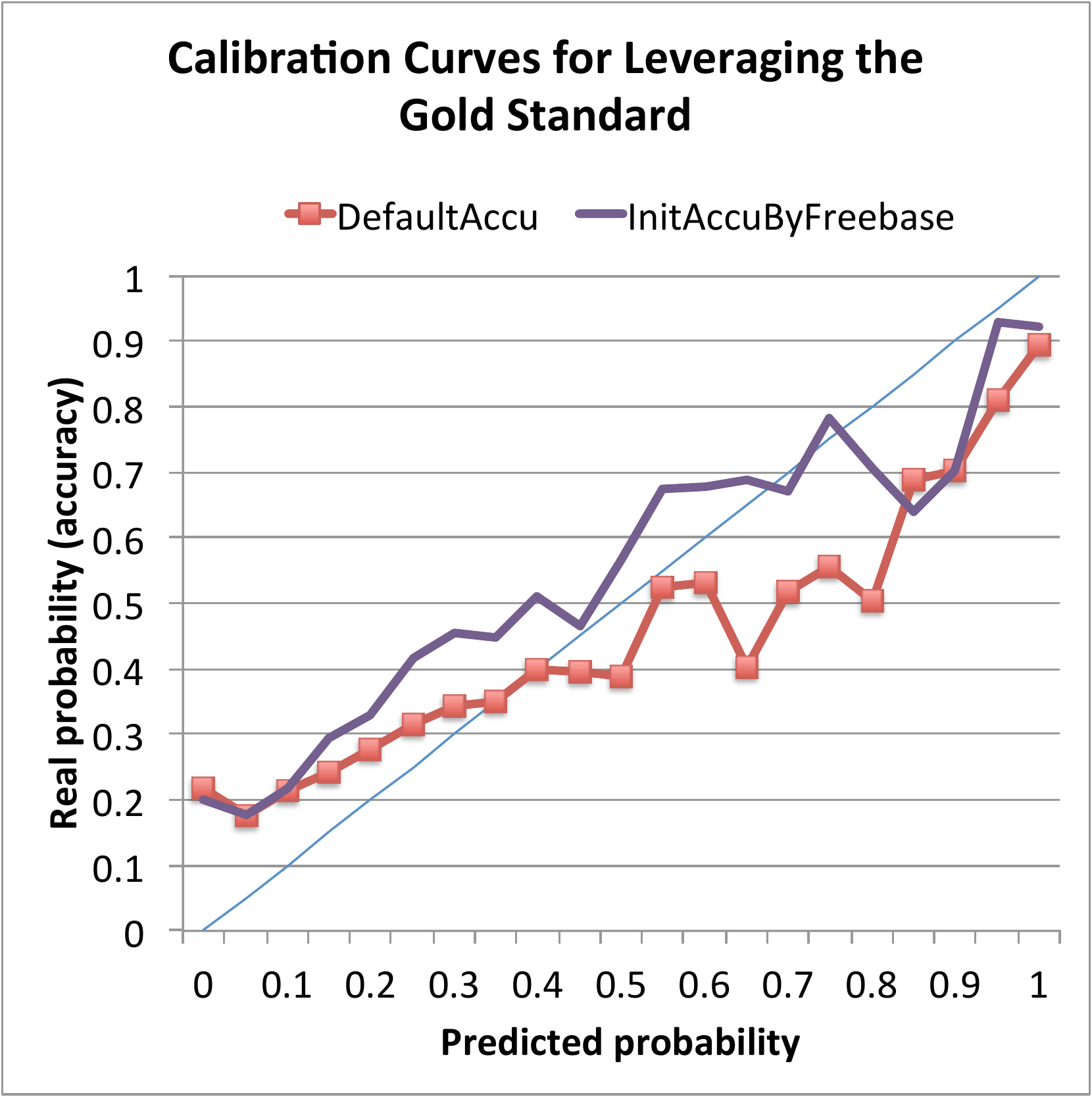}

\ \

\resizebox{\columnwidth}{!} {
\begin{tabular}{|c|c|c|c|}
\hline
 & Dev. & WDev. & AUC-PR \\
\hline
{\sc PopAccu} & .020 & .037 & .499 \\
{\sc InitAccu} (10\%) & .018 & .036 & .511 \\
{\sc InitAccu} (20\%) & .017 & .035 & .520 \\
{\sc InitAccu} (50\%) & .016 & .033 & .550 \\
{\sc InitAccu} (100\%) & {\bf .015} & {\bf .029} & {\bf .589} \\
\hline
\end{tabular}}
{\small
\caption{Leveraging the gold standard for initial accuracy computation
improves probability calibration.
\label{fig:kg}}
}
\end{minipage}
\hfill
\begin{minipage}[th]{.32\linewidth}
\centering
\includegraphics[scale=0.3]{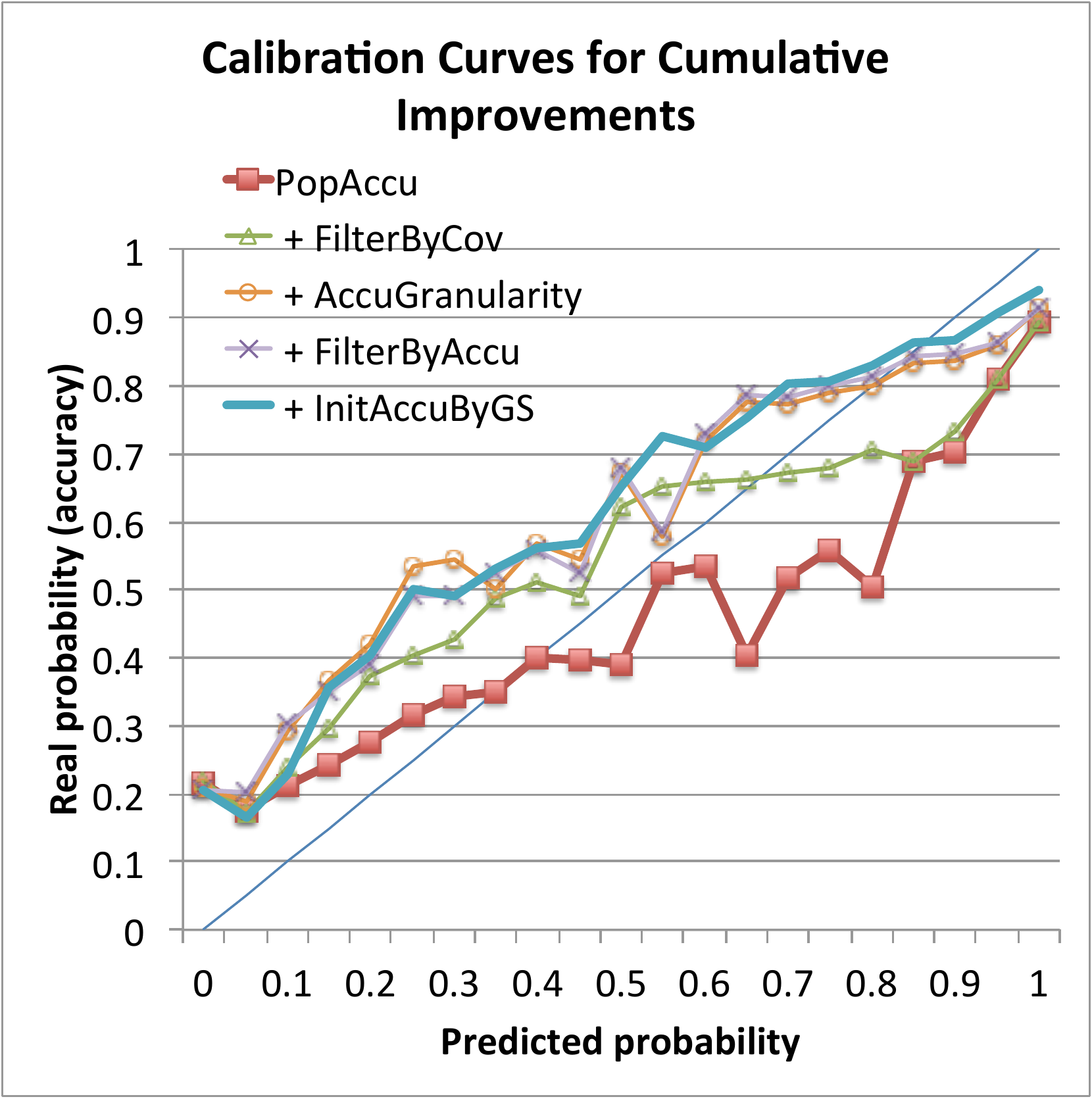}

\ \

\resizebox{\columnwidth}{!} {
\begin{tabular}{|c|c|c|c|}
\hline
 & Dev. & WDev. & AUC-PR \\
\hline
{\sc PopAccu} & .020 & .037 & .499 \\
+FilterByCov & .016 & .038 & .511 \\
+AccuGranularity & .023 & .036 & .544 \\
+FilterByAccu & .024 & .035 & .552 \\
+GoldStandard & {\bf .020} & {\bf .032} & {\bf .557} \\
\hline
\end{tabular}
}

{\small
\caption{The refinements altogether can significantly improve probability calibration
and AUC-PR.
\label{fig:together}}
}
\end{minipage}
\hfill
\begin{minipage}[th]{.32\linewidth}
\centering
\includegraphics[scale=0.3]{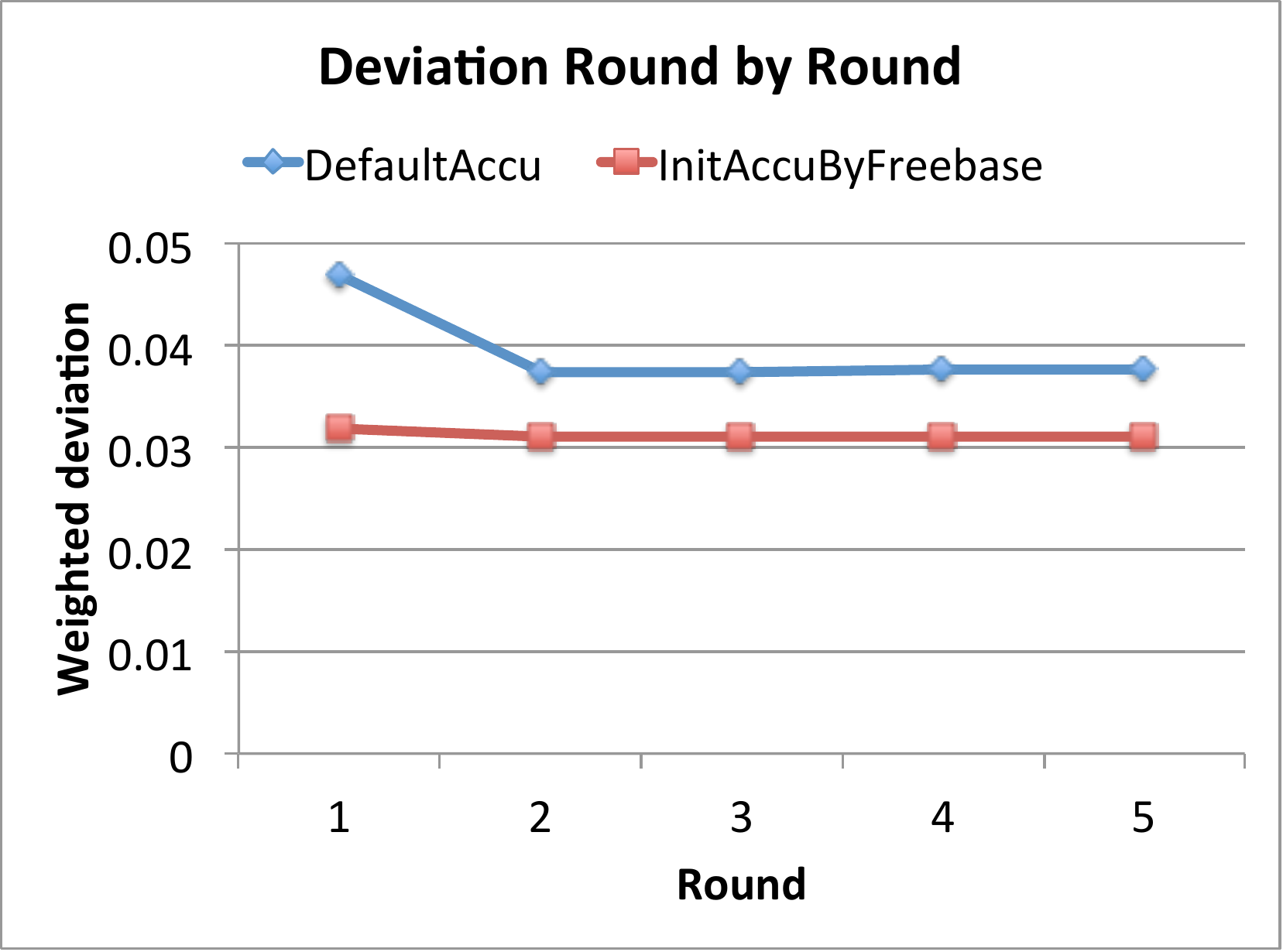}

\ \

\vspace{.4in}
{\small
\begin{tabular}{|c|c|c|c|}
\hline
 & Dev. & WDev. & AUC-PR \\
\hline
L=1M, R=5 & {\bf .020} & {\bf .037} & {\bf .499} \\
L=1K, R=5 & .020 & .037 & .499 \\
L=1M, R=25 & .019 & .038 & .497 \\
\hline
\end{tabular}
\caption{Weighted deviation vs number of iterations for two ways of
  initializing the provenance accuracies.
Sampling $L=1K$ triples in each reducer and
terminating in $R=5$ rounds does not hurt the results.
\label{fig:efficiency}}
}
\end{minipage}
\vspace{-.25in}
\end{figure*}
\subsection{Improving the existing methods}
\label{sec:refine}

We next describe a set of refinements that improve {\sc Accu} and {\sc PopAccu}.
We show our results for {\sc PopAccu}, since it has a lower
deviation than {\sc Accu}, but we have similar observations on {\sc Accu}.
For each refinement, we compare it with the default setting;
we show in the end the result of putting all refinements together.

\subsubsection{Granularity of provenances}
The basic models consider an (Extractor, URL) pair as a provenance.
We can vary the granularity. First, recall that half of the provenances
each contributes a single triple;
thus, we lack support data to evaluate the quality of such provenances.
If we consider the Web sources at a coarser level, the Website level
(\ie, the prefix of the URL till the first ``/''; for example,
replacing
``en.wikipedia.org/\\wiki/Data\_fusion'' with just ``en.wikipedia.org''),
we would have much more support data
for quality evaluation. So the first choice is between {\em page}-{\em level} and {\em site}-{\em level}.
Second, recall that we observe very different quality for different predicates;
thus, we may wish to evaluate quality for each predicate
instead of the whole Webpage or Website. So the second choice is between
{\em predicate}-{\em level} or all triples.
Third, recall that we observe very different quality of different patterns for
the same extractor; thus, we may wish to evaluate quality for each pattern
instead of the whole extractor. So the third choice is between
{\em pattern}-{\em level} and {\em extractor}-{\em level}.

Figure~\ref{fig:granularity} shows the calibration curves and various measures
for applying different granularities.
We observe that using site-level instead of source-level
has two effects. On the one hand, for triples whose predicted probability falls
in $[.65, 1]$, the predicted probability is much closer to the real probability;
this is because there are much fewer provenances for which we lack extracted triples
for accuracy evaluation. On the other hand, for triples whose predicted probability
falls between $[.05, .65)$, the predicted probability is farther away from
the real probability; this is partially because even triples extracted from
the same Website can still vary in quality, but now we cannot
distinguish them.

Using the predicate, as well as the extractor and the Website, further
improves the calibration (dragging the calibration curve closer
to the ideal diagonal line), since the finer-granularity enables
distinguishing quality of triples with different predicates from the same Website.
We note however that applying predicate-level granularity
together with
URL-level granularity would hurt the results, because we then have
even more provenances that contribute too few triples for effective accuracy evaluation.

Appending the pattern, in addition to site-level and predicate-level
granularity, further
improves the results, but only slightly. This is because some
extractors only
have a single pattern, and often a pattern extracts triples for just a single predicate.
Overall, comparing with the (Extractor, URL) granularity,
using (Extractor, site, predicate, pattern) reduces weighted deviation by
13\% and increases AUC-PR by 5\%.

\begin{figure*}
\vspace{-.1in}
\begin{minipage}[th]{.32\linewidth}
\centering
\includegraphics[scale=0.32]{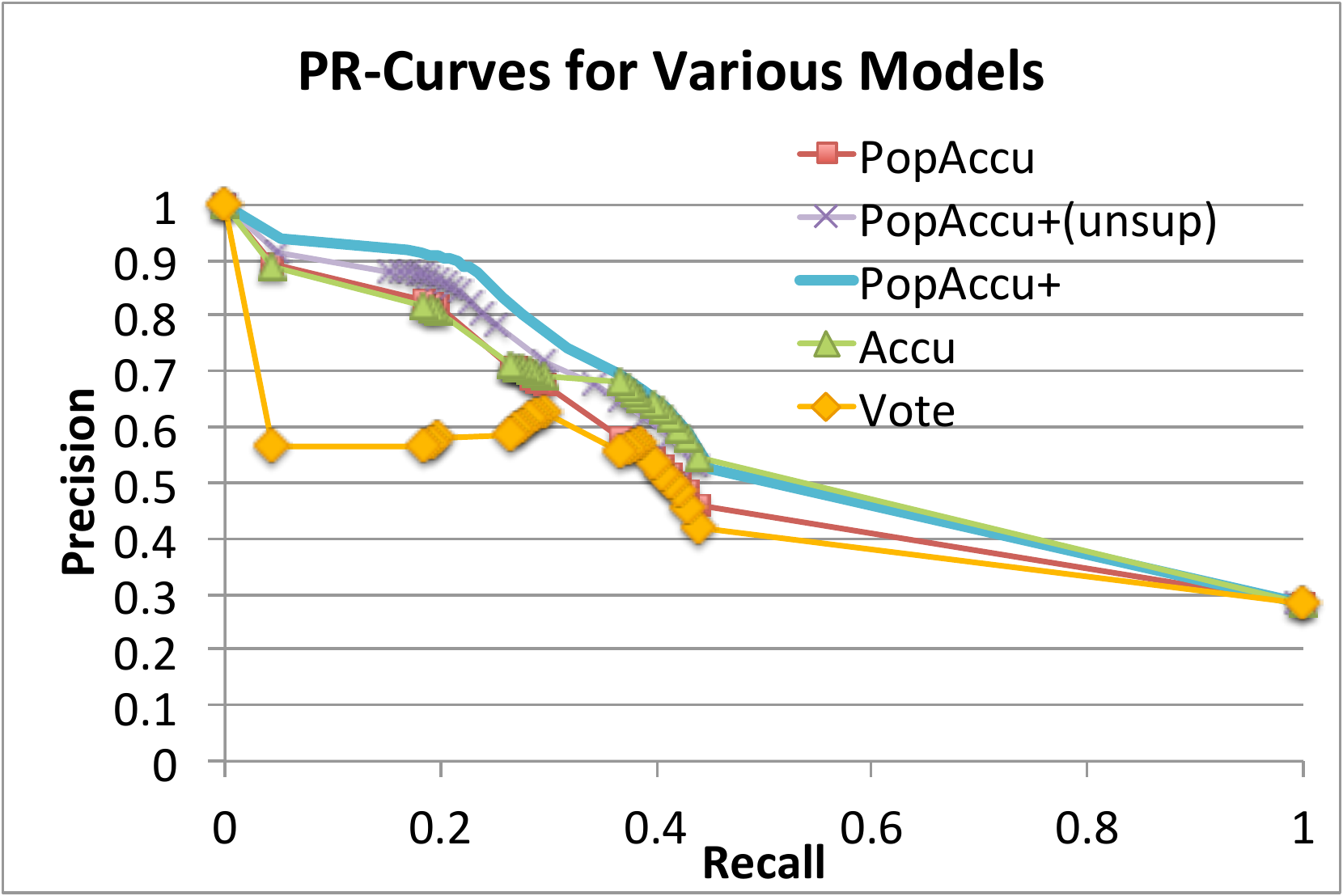}
\caption{The PR-curve for {\sc PopAccu+} has the best shape.
\label{fig:pr}}
\end{minipage}
\hfill
\begin{minipage}[th]{.32\linewidth}
\centering
\includegraphics[scale=0.32]{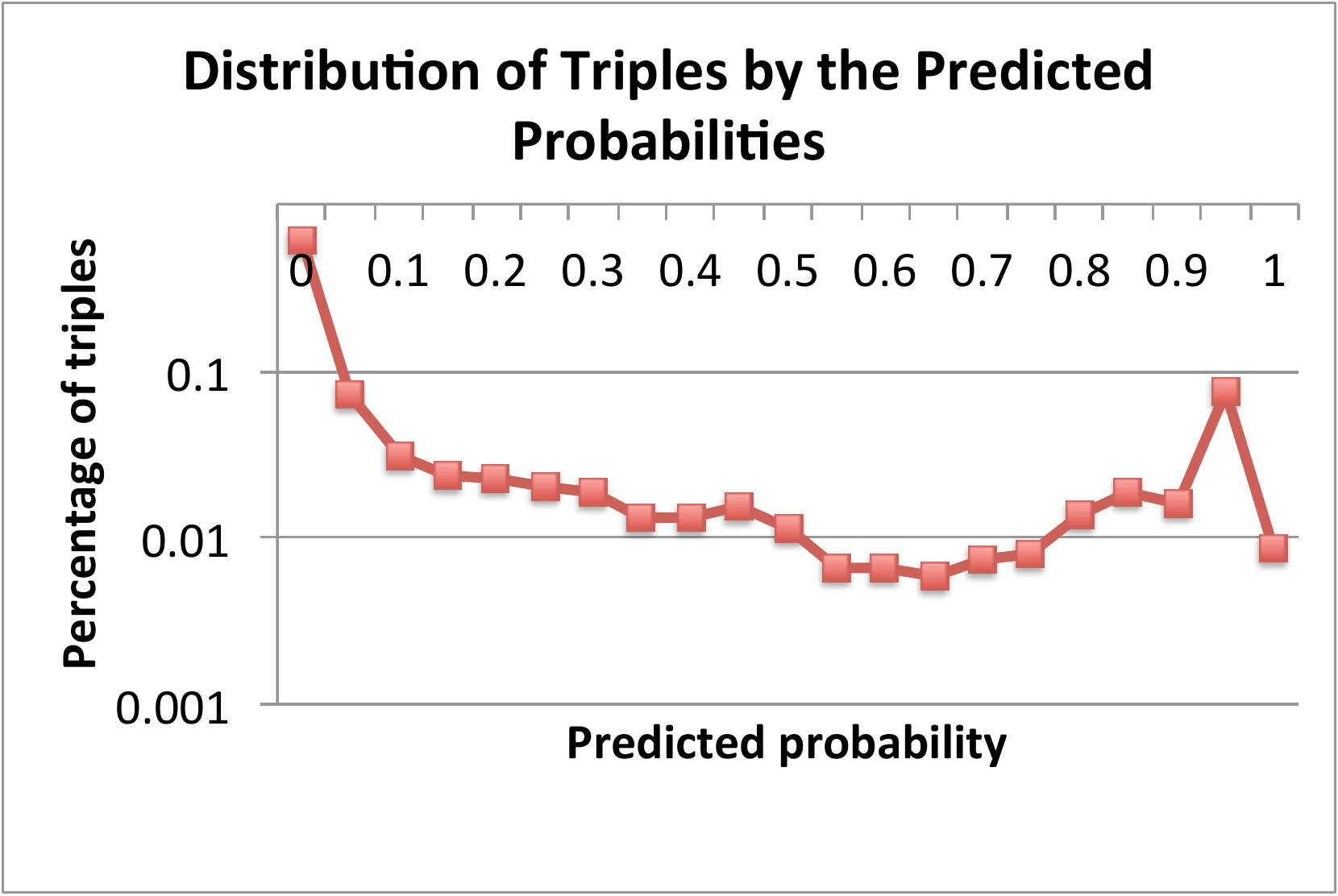}
\caption{
More than 70\% of the triples have a very low predicted probability.
\label{fig:dist}}
\end{minipage}
\hfill
\begin{minipage}[th]{.32\linewidth}
\centering
\includegraphics[scale=0.25]{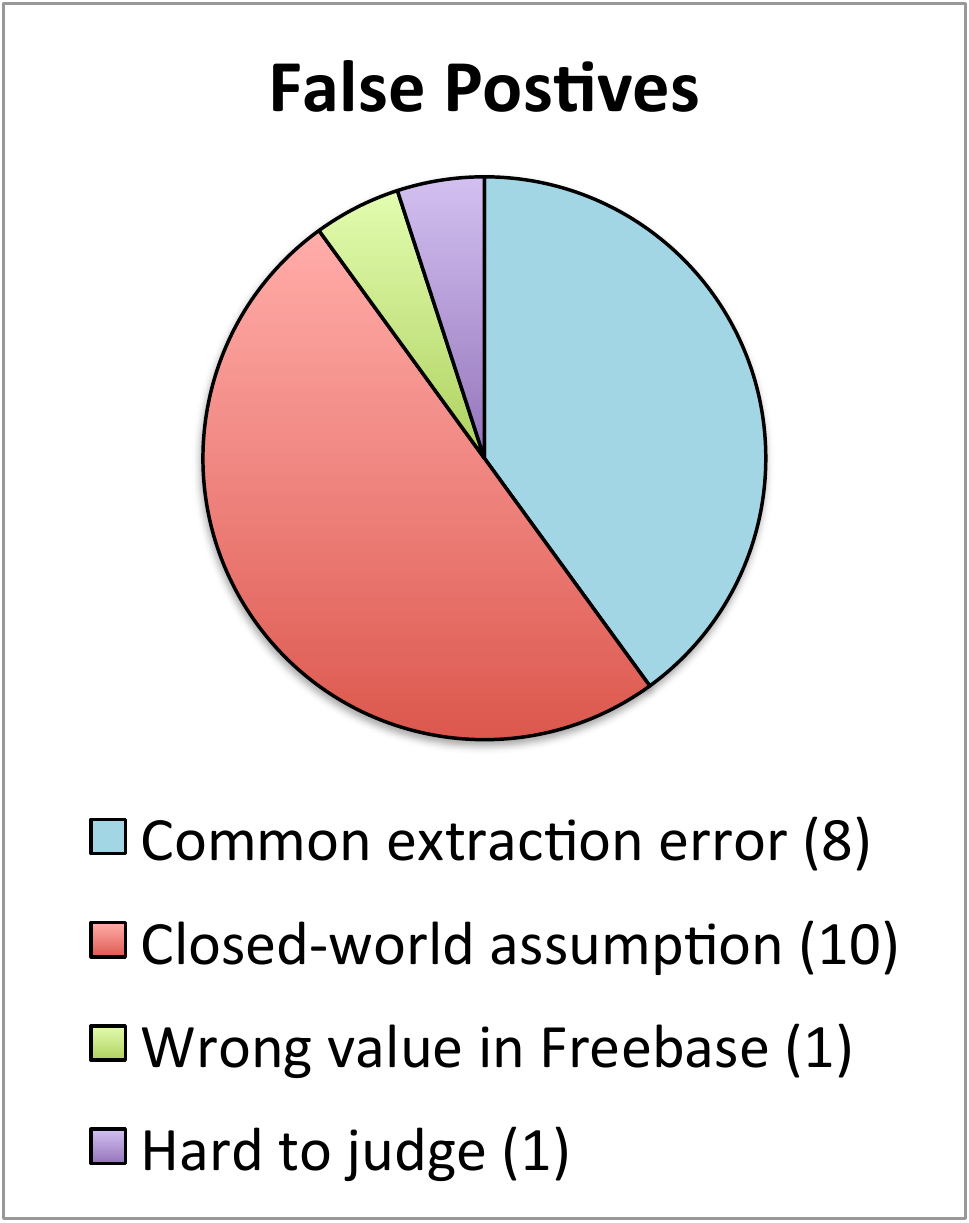}
\includegraphics[scale=0.25]{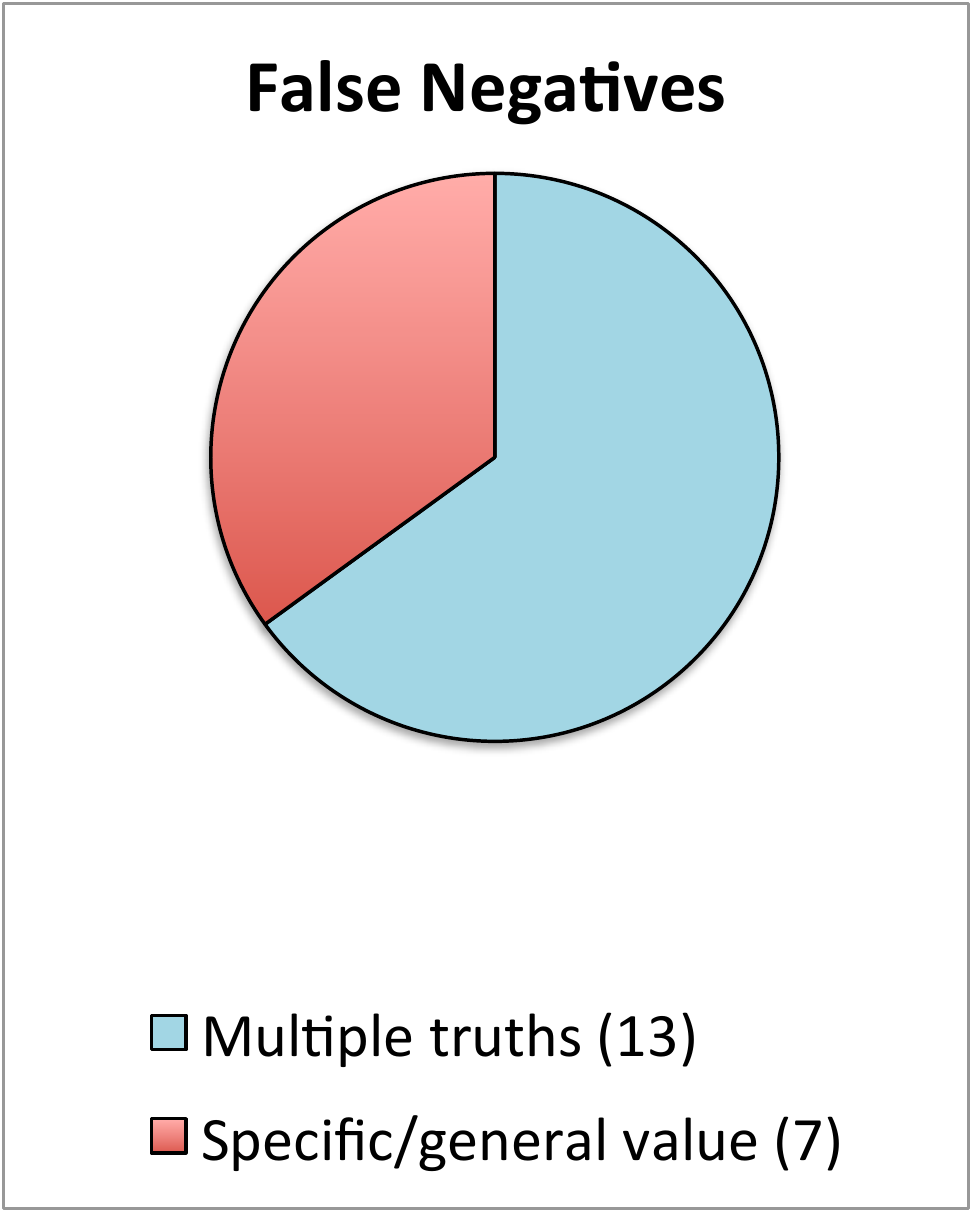}
\caption{Categorization of reasons for the errors by {\sc
    PopAccu+}
for a sample of 20 extractions.
\label{fig:pie}}
\end{minipage}
\vspace{-.15in}
\end{figure*}

\subsubsection{Provenance selection}
Recent work~\cite{DSS13} has shown that
ignoring low-quality sources can often obtain better results
in data fusion than considering every source.
Here we consider filtering provenances by two criteria ---
the coverage and the accuracy --- which we define below.

As we have seen, when a provenance contributes a single triple, we cannot effectively
evaluate its accuracy and this may hurt the probability we predict. We
revise our models in two ways to filter
provenances for which we cannot compute an accuracy other than the default accuracy.
(1) In the first round, when all provenances have
the same default accuracy, we compute triple probabilities for data items where
at least one triple is extracted more than once, and then re-evaluate accuracy
for each provenance. (2) In later rounds, we ignore provenances
for which we still use the default accuracy.
Figure~\ref{fig:filter} shows that this modification considerably smooths
the calibration curve; on the other hand, for 8.2\% of the triples,
we cannot predict a probability because the provenances are filtered.

We also filter provenances of low accuracy and revise our models as follows.
(1) Given a threshold on accuracy $\theta$, we ignore a provenance if its
accuracy is below $\theta$. (2) For certain
data items we may lose all provenances so cannot predict the probability for any
triple; as a compensation, for such cases we set the probability
of each triple as the average accuracy of its provenances.
In Figure~\ref{fig:filter}
we observe that setting $\theta=.1$ would significantly improve the calibration curve,
reducing the weighted deviation by 5\%. Setting a higher $\theta$ can continue increasing
AUC-PR, but may not improve the calibration curve any more,
since for more and more triples we compute the probability
as the average accuracy of the provenances instead of applying the principled Bayesian analysis.
Starting from $\theta > .5$, even the AUC-PR measure decreases.

\subsubsection{Leveraging the gold standard}
There are already high-quality knowledge bases such as {\em Freebase}
and we can leverage them in predicting
the correctness of newly extracted knowledge triples. 
Recall that we initialize the accuracy of a provenance as a default value .8;
instead, we compare its data with the gold standard,
created using {\em Freebase} under the local closed-world assumption,
and initialize the accuracy as the percentage of true triples.

Figure~\ref{fig:kg} shows that this change
can improve the results a lot: it decreases the weighted deviation by 21\% and increases
AUC-PR by 18\%. In addition, when we sample only a portion of the gold standard,
we observe that the higher sample rate, the better results; we obtain the best results
when we use the full gold standard.

\eat{
However, the second approach fails. When we simply considering the gold standard
as an extra source, we observe a calibration curve with very different shape.
On the one hand, if a false triple is provided by a large number of sources,
the strength of the new source, even though with a high probability,
is not big enough to make a big difference. On other hand, for data items where
{\em Freebase} knows multiple true triples, {\sc PopAccu} computes a lower probability for each of them,
causing over-estimate in many cases.
Moreover, after we combine the first and the second approaches, while we can significantly
improve AUC-PR, the calibration of the predicted probabilities is significantly worse.
Now for each true triple, with both the extra sources and likely higher-accuracy provenances,
we are more likely to compute a high probability for it than the conflicting false triples.
However, we tend to compute very similar probabilities for all the true triples
on a non-functional predicate, and these probabilities are likely to be low since
they add up to 1.
}

\subsubsection{Putting them all together}
We now apply all the refinements and make the following changes one by one:
I. filter provenances by coverage; II. change provenance
granularity to (Extractor, pattern, site, predicate); III. filter provenances by accuracy
(here we observe the best results when we set $\theta=.5$); and IV. initialize
source accuracy by the gold standard.
After the first three changes the algorithm is still unsupervised,
which we call {\sc PopAccu}$^+_{unsup}$; with the fourth change the algorithm
becomes semi-supervised, which we call {\sc PopAccu+}.

Figure~\ref{fig:together} shows the calibration curve and the measures
for each method,
and
Figure~\ref{fig:pr} shows the corresponding PR-curves.
The changes altogether reduce the weighted deviation by 13\%,
and increases the AUC-PR by 12\%.
We observe that Change I can smoothen the calibration curve a lot.
Change II not only improves the calibration, but also increases
the percentage of the triples for which we predict probabilities from 91.8\% to 99.4\%.
Change III further improves the results, but very slightly.
Change IV also further improves the results, mainly for triples of very high ($[.85, 1]$)
or very low ($[0, .15)$) predicted probabilities, and smoothes out for triples with
predicted probabilities in the range of $[.45, .6)$.
However, with the improvements from Changes I-III, the help from Change IV is limited.
We note that only applying Change IV can obtain a lower deviation and a higher
AUC-PR, as it overfits when we have fine-grained provenances;
however, the calibration curve is not smooth because of the overfitting
(see Figure~\ref{fig:kg}).

Figure~\ref{fig:dist} shows the distribution of the predicted probabilities
for {\sc PopAccu+}. We observe that most of the triples have very high
or very low probabilities: 70\% triples are predicted with
a probability of lower than 0.1, while 10\% triples are predicted with
a probability of over 0.9.

\subsubsection{Speeding up execution}
\eat{
Since we ran our experiments on the Google Borg system, the execution time would
be affected by scheduling and by the workload from other jobs at the same time.
{\sc Basic} has one single iteration and typically finished in 50-60 minutes.
For {\sc Accu} and {\sc PopAccu},
on average each iteration (including Stage I and Stage II) took 75 minutes;
the overall running time, including in addition pre-processing,
result outputting (Stage III), scheduling, etc., can take 7-10 hours.
}

Finally, we evaluate the choices we have made in speeding up execution.
Recall that we sample $L=1M$ triples for each data item
in triple-probability computation and for each provenance in accuracy evaluation.
Not applying such sampling would cause out-of-memory errors, while
sampling $L=1K$ triples actually leads to very similar performance measures
(Figure~\ref{fig:efficiency}).

Recall also that we force termination after $R=5$ rounds. Figure~\ref{fig:efficiency}
shows that if we set default accuracy initially for each provenance, the predicted triple
probabilities would change a lot from the first round to the second,
but stay fairly stable afterwards.
If we initialize provenance accuracy by the gold standard,
even the difference from the first two rounds is small.
Terminating after $R=25$ rounds obtains very similar
performance measures.

\begin{figure*}
\vspace{-.1in}
\begin{minipage}[th]{.32\linewidth}
\centering
\includegraphics[scale=0.3]{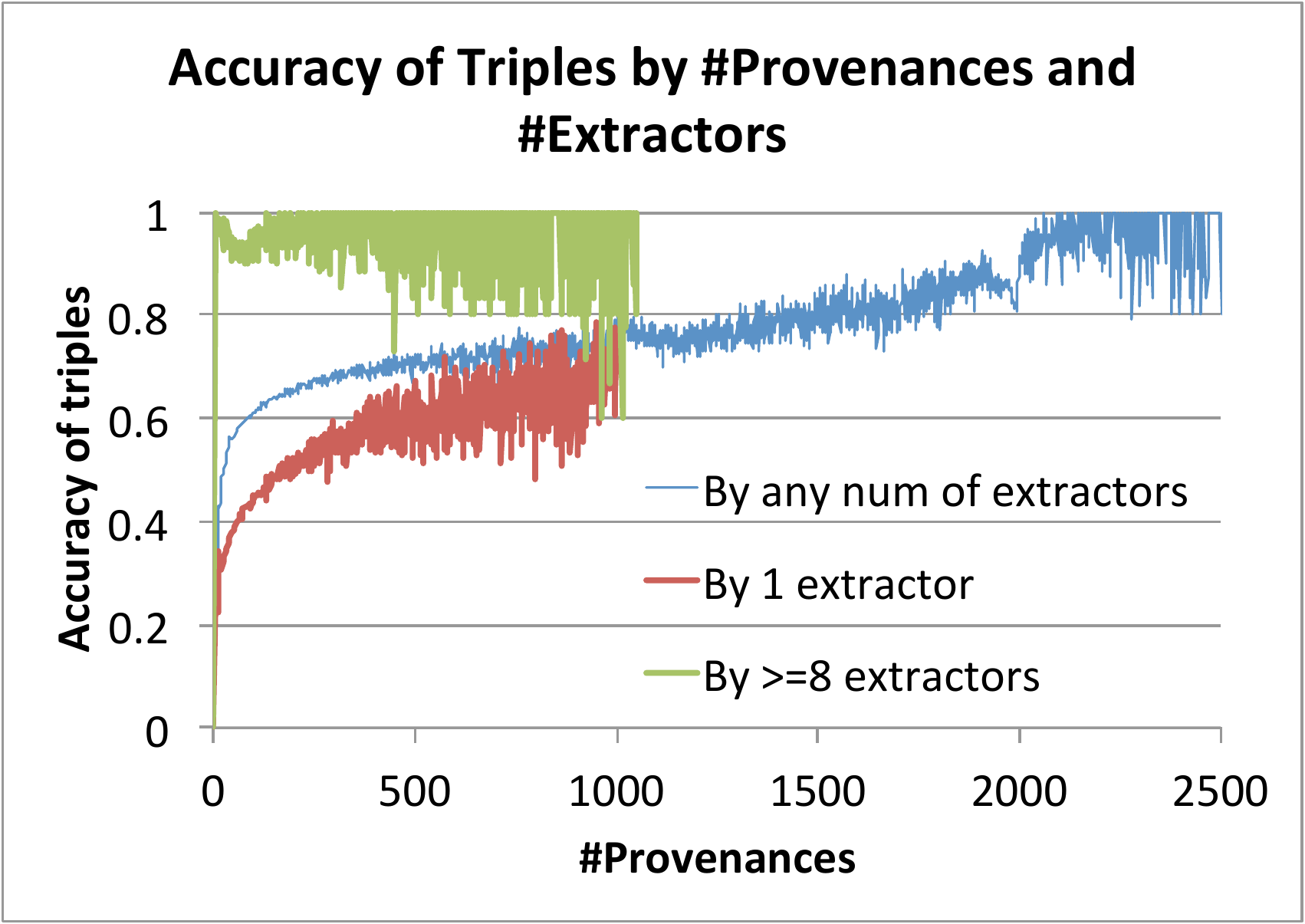}
\caption{
Fixing \#provenances, triples from more extractors
are more likely to be true.
\label{fig:accuByProv}}
\end{minipage}
\hfill
\begin{minipage}[th]{.32\linewidth}
\centering
\includegraphics[scale=0.3]{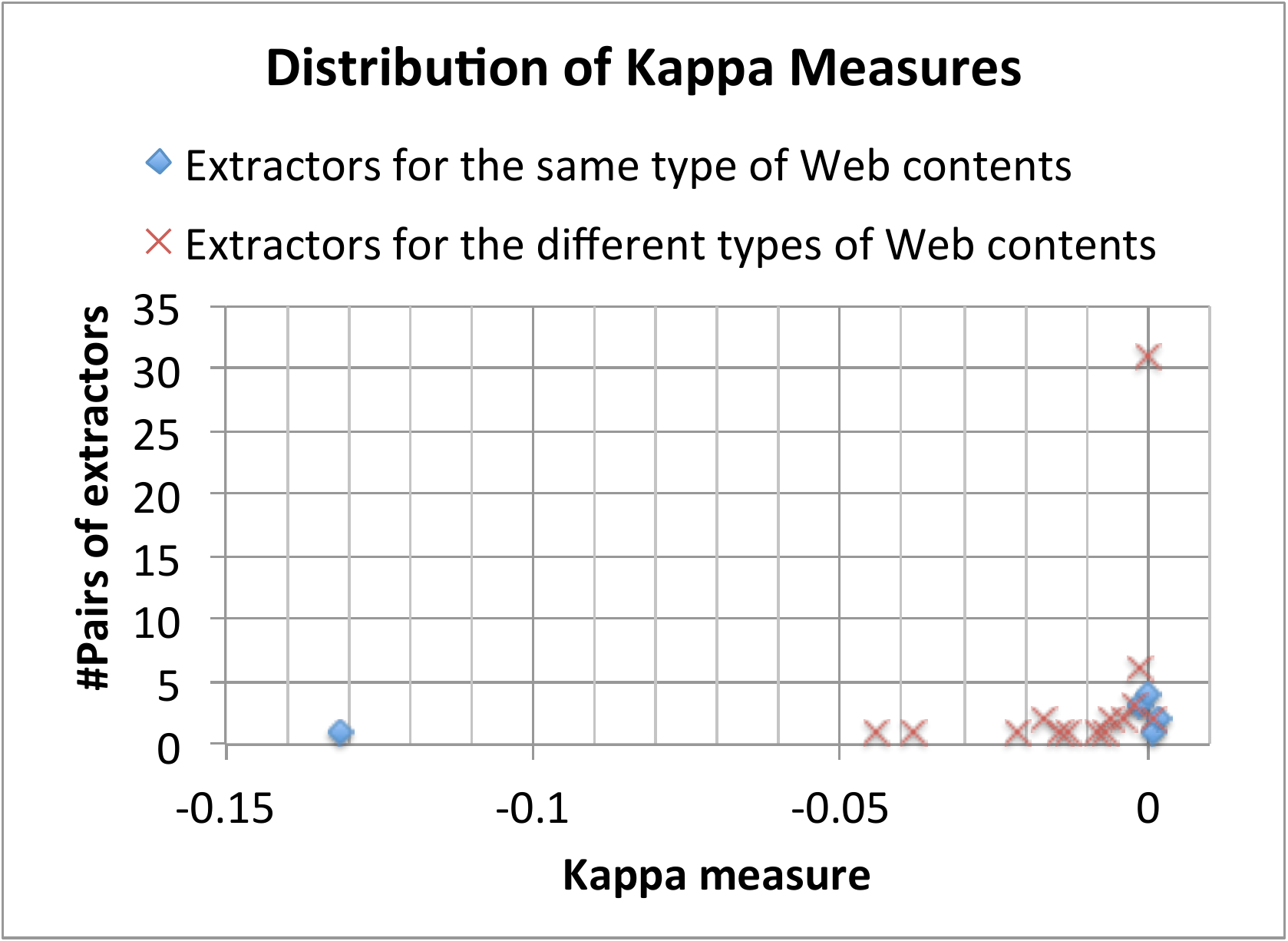}
\caption{The Kappa measure indicates that a lot of extractors
are anti-correlated.
\label{fig:kappa}}
\end{minipage}
\hfill
\begin{minipage}[th]{.32\linewidth}
\centering
\includegraphics[scale=0.3]{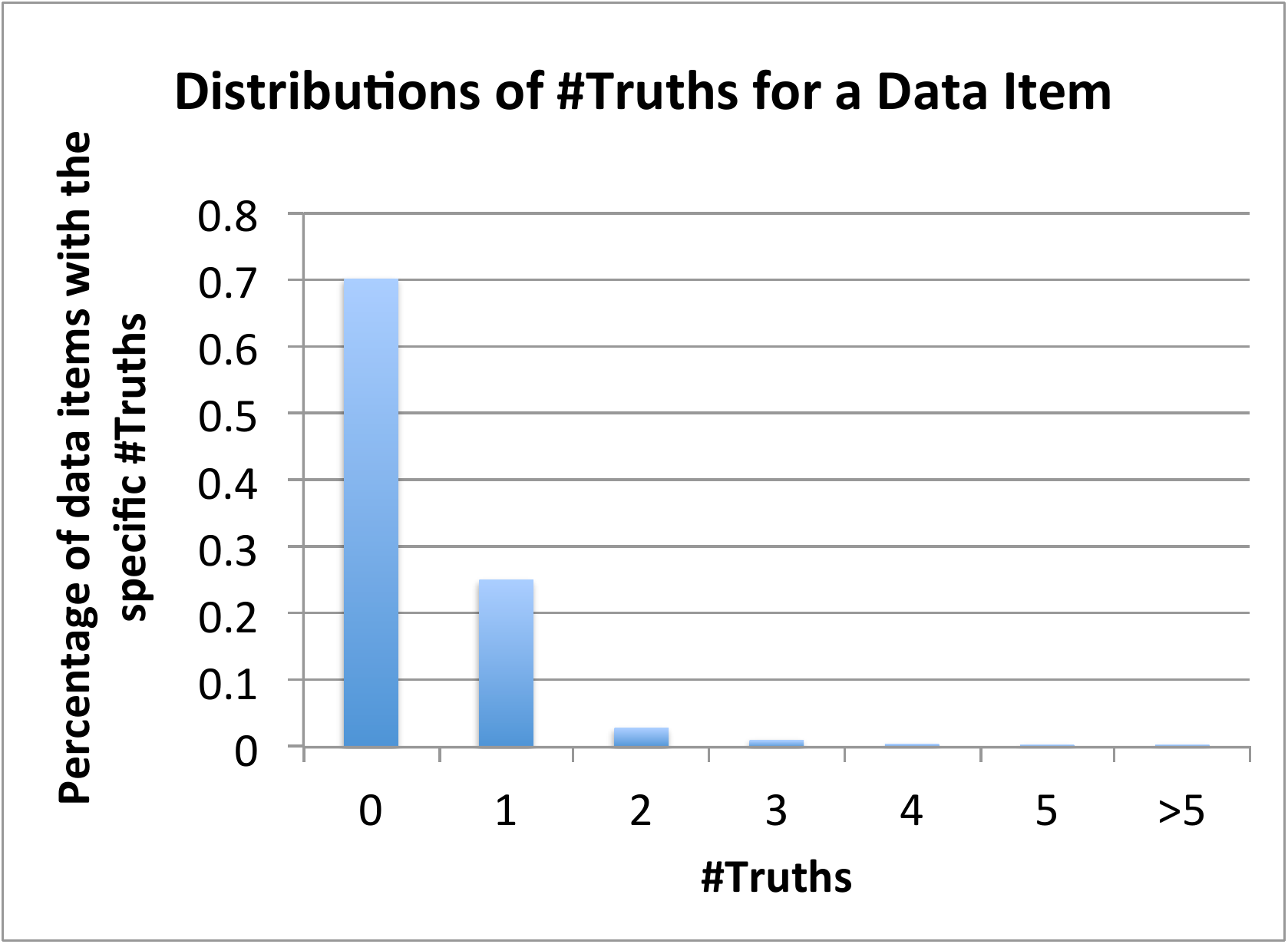}
\caption{We know multiple true triples in the gold standard
for very few data items. \label{fig:nTruths}}
\end{minipage}
\vspace{-.1in}
\end{figure*}

\subsection{Error analysis}
\label{sec:err}
Although {\sc PopAccu+} improves a lot over {\sc PopAccu} and has reasonably
well calibrated curves, there is still a lot of space for improvement.
Among the triples that {\sc PopAccu+} predicts to be true with a fairly high confidence
(predicting a probability above 0.9),
the accuracy is 0.94, so there are still quite some false positives.
On the other hand, recall that 30\% of the extracted triples are correct,
but {\sc PopAccu+} only predicts a probability over 0.5 for
18\% of them, so there are still a lot of false negatives.

To gain insights on how we can further improve knowledge fusion,
we next take a closer look at the errors made by {\sc PopAccu+}.
We randomly selected 20 false positives (\ie, triples for which
we predict a probability of 1.0 but the gold standard
says are false), and 20 false negatives (\ie, triples for which we predict
a probability of 0.0 but the gold standard says are true). We manually checked
the Webpages from which each triple is extracted and decided the reason for
the ``mistakes'', summarized in Figure~\ref{fig:pie}.

Of the 20 false positives, 8 (40\%) are caused by common extraction errors
made by one or two extractors on a lot of Webpages:
3 are triple identification errors,
3 are entity linkage errors, and 2 are predicate linkage errors.
Of the remaining false positives, 10 (50\%) actually are not errors,
but are classified as such due to the local closed world assumption.
Specifically,
for 5 triples {\sc PopAccu+} finds additional correct values that were
not in {\em Freebase};
for 3 triples {\sc PopAccu+} chooses a more specific (but correct) value
(e.g., ``{\em New York City}''' is a more specific value than ``{\em USA}'');
and for 2 triples {\sc PopAccu+} chooses a more general (but correct)
value.
One false positive is due to {\em Freebase} having an
obviously incorrect value.
And finally, one false positive is hard to judge: the extracted value
is mentioned on a lot of Webpages but contradicts the value in {\em Freebase};
both values seem plausible.

There are two major reasons for false negatives.
For 13 (65\%) triples there are multiple truths for the associated data item;
{\sc PopAccu+} computes a high probability for only one of them because of
its single-truth assumption. For 7 (35\%) triples the values are hierarchical;
{\sc PopAccu+} chooses one value under the single-truth assumption
but predicts a low probability for a more specific or a more general value.
We note that for some false positives, presumably there can be a false
negative if the truth is extracted,
since the probabilities for all triples associated with the same
data item sum up to 1. None of them appears in our sample though,
indicating that the percentage is low.

\section{Future directions}
\label{sec:improvement}
Although our experimental results suggest that small modifications to
existing DF methods can lead to reasonable results for the KF task, there is still
much room for improvement. However, improving the quality will require
more radical changes to the basic assumptions made by DF methods.
We sketch some of these future directions in this section.


\vspace{-.05in}
\subsubsection*{1. Distinguishing mistakes from extractors and from sources}
One key assumption that DF techniques make is that a value provided by
a large number of data sources is more likely to be true, unless copying
exists. This assumption is verified on Deep Web
data~\cite{LDL+12}, but breaks on extracted knowledge when we consider
an (Extractor, URL) pair (or provenances of different granularities) as a source.
Recall that among the 20 randomly selected false positives in our error analysis,
8 of them are genuine errors and they all come from
common extraction errors by one or two extractors. {\sc PopAccu+} fails to
detect such extraction errors when they are prevalent.

To illustrate this, we plot in Figure~\ref{fig:accuByProv} the accuracy of
triples by the number of provenances. Because of possible common extraction errors,
having a lot of provenances is not necessarily a strong signal for correctness:
the triples with 100 provenances have an accuracy of only 0.6,
and the triples with 1000 provenances have an accuracy of only 0.75.
However, for triples with the same number of provenances,
those extracted by at least 8 extractors have a much higher accuracy
(on average 70\% higher) than those extracted by a single extractor.
This important signal, unfortunately, is buried when we
simply represent the provenance as the cross product of
the Web source and the extractor.

A better approach would be to distinguish mistakes made by extractors
and erroneous information provided by Web sources. This would enable us
to evaluate the quality of the sources and the quality of the
extractors independently. Then, we could
identify possible mistakes by the same extractor on many
different Web sources, and avoid being biased by a false triple
provided by only a couple of sources but extracted by many different extractors.

\begin{figure*}
\vspace{-.1in}
\begin{minipage}[th]{.65\linewidth}
\centering
\includegraphics[scale=0.3]{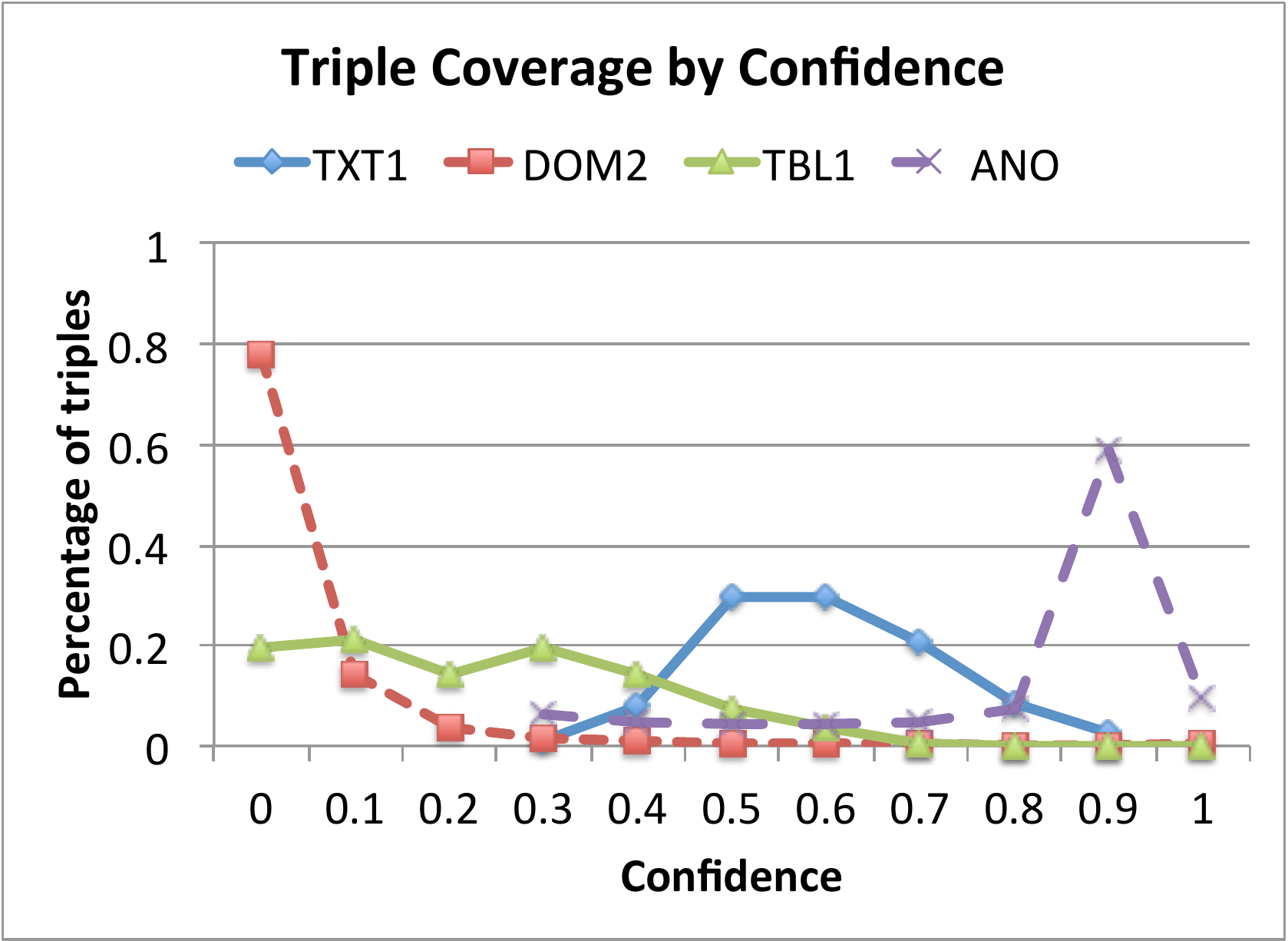}
\hfill
\includegraphics[scale=0.3]{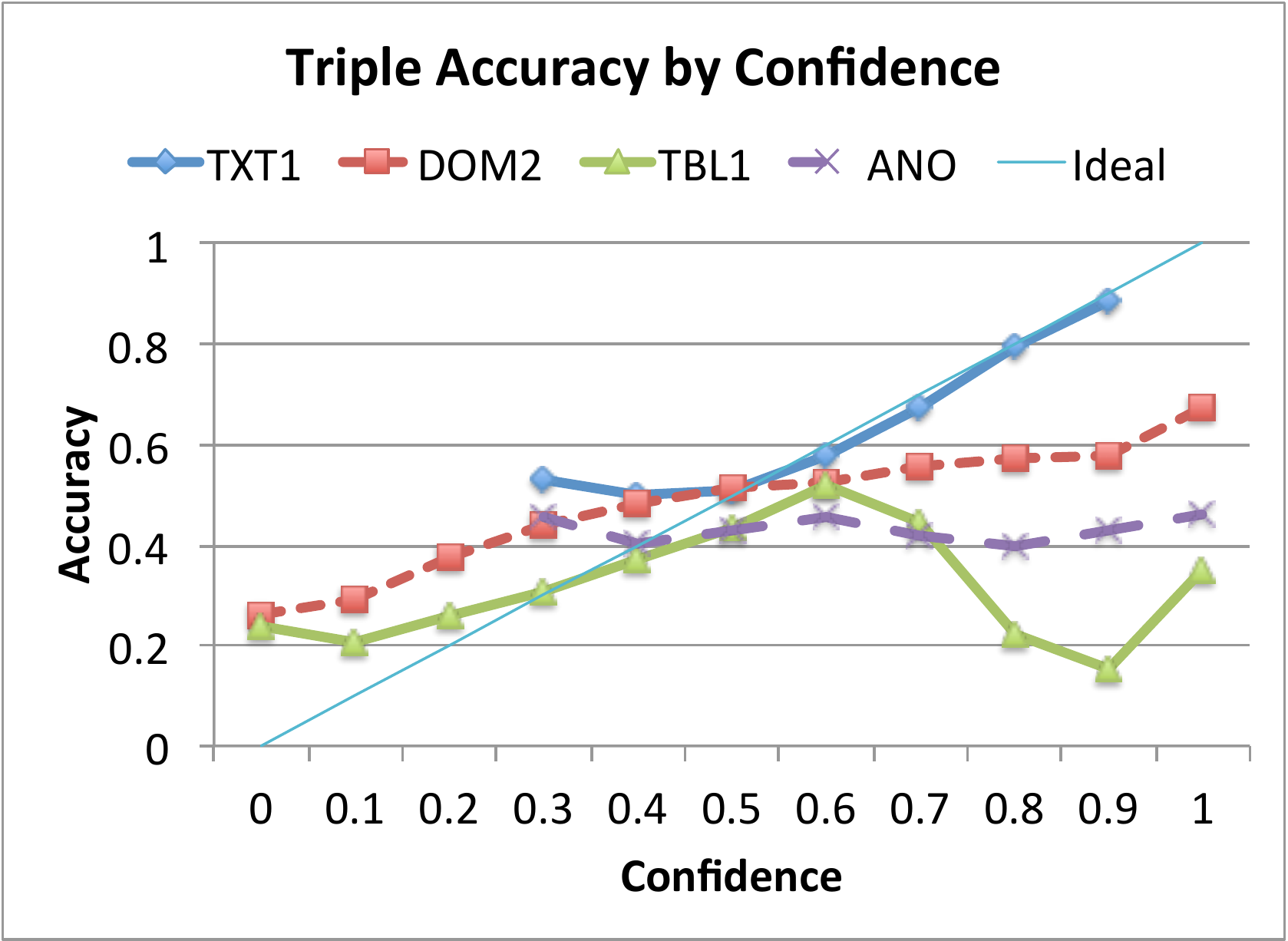}
\caption{
Coverage and accuracy by extraction confidence for four example
extractors show big difference among the extractors.
\label{fig:conf}}
\end{minipage}
\hfill
\begin{minipage}[th]{.32\linewidth}
\centering
\includegraphics[scale=0.3]{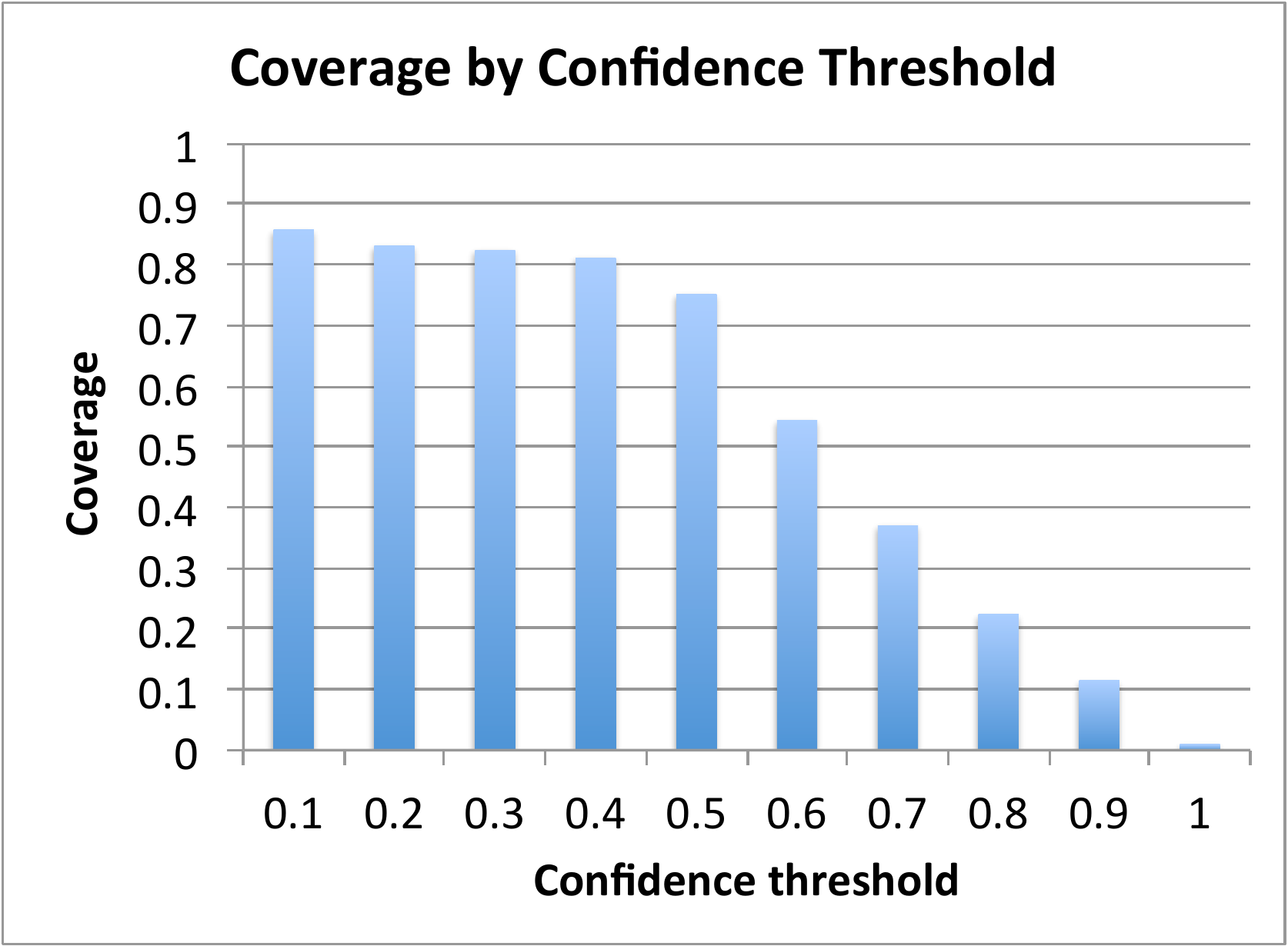}
\caption{Filtering triples by confidence can significantly drop coverage.
\label{fig:confThreshold}}
\end{minipage}
\vspace{-.2in}
\end{figure*}

\vspace{-.05in}
\subsubsection*{2. Identifying complex correlations between extractors and
between sources}
Both {\sc Accu} and {\sc PopAccu} assume independence between data sources.
In our adaptations a data source is an (Extractor, URL) pair, so obviously
the sources are {\em not} independent.
Existing DF works have discussed how to detect copying
between data sources by finding common mistakes, since independent sources
are less likely to make a lot of common mistakes~\cite{BCM+10, DBH+10a, DBS09a, DBS09b, LDOS11}.
Such techniques can fall short in KF for two reasons. First and most importantly,
we have billions of Web sources, so the proposed techniques that
reason about every pair of sources do not scale.
Second, for Web sources we have only {\em extracted} data instead of knowing
the truly {\em provided} data, so the proposed models can wrongly consider
common extraction errors as evidence of copying. 

In addition, for extractors
the relationships can be much richer: instead of copying,
there can be correlation or anti-correlation.
To illustrate such rich relationships, we plot in Figure~\ref{fig:kappa}
the distribution of the {\em Kappa measure}~\cite{kappa} of every pair of extractors
designed for the same type of Web contents (\eg, TXT, DOM) and those for different types of
Web contents. The Kappa measure of two sets of extracted triples ${\bf T}_1$
and ${\bf T}_2$ with respect to the overall set ${\bf KB}$ is defined as follows.

\vspace{-.1in}
\begin{equation}
\kappa = {|{\bf T}_1 \cap {\bf T}_2|\cdot|{\bf KB}| - |{\bf T}_1|\cdot|{\bf T}_2| \over |{\bf KB}|^2 - |{\bf T}_1|\cdot|{\bf T}_2|}
\end{equation}

\noindent
The Kappa measure is considered as a more robust measure than merely measuring
the intersection, as it takes into account the intersection that can happen
even in case of independence.
A positive Kappa measure indicates positive correlation;
a negative one indicates negative correlation;
and one close to 0 indicates independence.
Among the 66 pairs of extractors, 53\% of them are independent.
Five pairs of sources are positively correlated (but the kappa measures are
very close to 0), as they apply the same extraction techniques
(sometimes only differ in parameter settings) or investigate the same type of Web contents.
We observe negative correlation on 40\% of the pairs;
they are often caused by considering different types of Web contents,
but sometimes even extractors on the same type of Web contents can be highly
anti-correlated when they apply different techniques.
Finally, we point out that since a lot of extractors employ the same
entity linkage components, they may make common linkage mistakes,
even if the Kappa measure indicates independence or anti-correlation.
Finding general correlations has been studied in~\cite{PDD+14, QAH+13},
but again, it is not clear how to scale the reasoning for billions of sources.

We wish to scale up existing methods
such that we can reason about copying between Web sources
(recall that for one false positive in our
error analysis we are not sure if a wrong fact has spread out)
and the rich correlations between extractors. 

\begin{table}
\centering
{\small
\caption{The majority of the knowledge triples have non-functional predicates.
\label{tbl:single-multi}}
\vspace{-.1in}
\begin{tabular}{|c|c|c|c|c|}
\hline
Type & Predicates & Data items & Triples & Accuracy\\
\hline
Functional & 28\% & 24\% & 32\% & 0.18 \\
Non-functional & 72\% & 76\% & 68\% & 0.25 \\
\hline
\end{tabular}
}
\vspace{-.2in}
\end{table}

\vspace{-.05in}
\subsubsection*{3. Handling non-functional predicates properly}
One big limitation of the DF models implemented in this paper
is that they all assume predicates being functional;
that is, for each data item the probabilities of different triples
add up to 1. This assumption is invalid for the majority of the data items
in our extracted knowledge:
as shown in Table~\ref{tbl:single-multi}, 72\% of the predicates are actually
non-functional and 76\% of the data items have non-functional predicates.
This assumption is responsible for 65\% of the false negatives
according to our error analysis.

Despite the invalidity of our assumption, the performance of our methods is not too bad:
as shown in Figure~\ref{fig:together}, when {\sc PopAccu+} predicts
a probability below 0.1, the real accuracy is  0.2, showing that the functionality
assumption does not hurt the results that much.
To understand why, we plot in Figure~\ref{fig:nTruths} the distribution
of the number of truths in the gold standard for each data item.
We observe that for 70\% of data items, all extracted triples are false;
for 25\% data items, a single extracted triple is correct;
and for only 3\% data items are two extracted triples correct.
In other words, we do not miss any true triple for 95\% of the data items.
(We admit that the gold standard may miss true triples because of the
local closed world assumption; we discuss that issue below.)

Nevertheless, we should develop methods that can handle functional
and non-functional predicates.
Zhao et al.~\cite{ZRHG12} proposed a graphical model that can predict truthfulness
when there are multiple truths. Instead of reasoning about source accuracy,
the model considers {\em sensitivity} (\ie, recall) and {\em specificity} of each source when deciding
value correctness. An interesting direction for future work is to
devise a way to scale up this approach for KF.

An even better approach, inspired by Figure~\ref{fig:nTruths},
is to learn the degree  of functionality
for each predicate (\ie, the expected number of values), and to
leverage this when performing fusion.
For example, most people only have a single spouse, but most actors
participate in many movies; thus the spouse predicate is ``nearly
functional'' whereas the acted-in predicate is highly
``non-functional''.


\subsubsection*{4. Considering hierarchical value spaces}

The current DF implementations treat the objects as categorical values;
that is, two object values are considered completely different if they do not match exactly.
There are two limitations to this. First,  values can be
hierarchically structured: for example,
{\em North America}-{\em USA}-{\em CA}-{\em San Francisco County}-{\em San Francisco}
forms a chain in the location hierarchy. Because of such value hierarchy,
even for data items with functional predicates there can be multiple
truths (\eg, the triples {\em (Steve Jobs, birth\_place, USA)} and 
{\em (Steve Jobs, birth\_place, California)} can both be true).
Ignoring value hierarchy accounts for 35\% of the false negatives in our results.
Furthermore, it misses the evidence hidden in the hierarchy. For example,
a triple with object {\em CA} partially {\em supports} that {\em San Francisco} is
a true object, but does not support {\em NYC} as a truth; on the other hand,
if several cities in {\em CA} are provided as conflicting values for a data item,
although we may predict a low probability for each of these cities,
we may predict a high probability for {\em CA}.

Second,  values can be similar to each other; for example, {\em 8849} and {\em 8850}
are similar in their numerical value, 
and the entity referring to {\em George Bush} and that referring to
{\em George W. Bush} are similar in their representations.
A triple with a particular object presumably also partially supports
a similar object.

We are not aware of any existing work that considers value hierarchy.
Previous works such as~\cite{DBS09a, LDL+12, PR11, YHY07}
have shown that considering value similarity can improve the results;
however, they all focus on similarity of values for strings, numbers, etc.
We need a strategy that can reason about the hierarchy and similarity of {\em entities},
where the information is presented by (sometimes wrong) triples
in the extracted knowledge.

\vspace{-.05in}
\subsubsection*{5. Leveraging confidence of extractions}

All of our implemented models assume deterministic data and treat all extracted
triples equally. However, for 99.5\% of the extracted triples,
the extractor provides a confidence indicating how likely the extraction is correct.
Note that a confidence is different from a probability in that
it may not be calibrated, making it challenging in incorporating them in
knowledge fusion.

To illustrate this, we plot in Figure~\ref{fig:conf}
the coverage and the accuracy of the triples
by their confidence from four selected extractors. We observe a big
difference in the distribution of confidence scores across extractors:
some extractors (DOM2 and ANO) tend to assign confidence close to 0 or 1,
while some extractors (TXT1) tend to assign confidence close to 0.5.
We also observe a big difference in the distribution of accuracy
across extractors:
some extractors (TXT1 and DOM2) are effective in computing confidence--the accuracy and the
confidence are in general correlated;
some extractors (ANO) do not give very useful confidence--the accuracy of the triples
stays similar when the confidence increases;
some extractors (TBL) may be very bad at predicting confidence--the peak of the accuracy
occurs when the confidence is medium.

\eat{
As we have shown in Figure~\ref{fig:conf}, different extractors may have completely
different distributions for confidence assignment and the effectiveness varies as well.
But there is one commonality: when a triple has a very low confidence, the likelihood
of it being correct is also low.
Our current model does not take into account extraction confidence at all but
the results are quite well. However, we would expect considering extraction confidence
can further improve the results. {\em How many false positives have low confidence?}
}

One obvious solution is to filter triples by the confidence; however, as shown
in Figure~\ref{fig:confThreshold}, even using a threshold as low as 0.1, we
already lose 15\% of the extracted triples.
Pasternack and Roth~\cite{PR11} considered confidence provided by the sources in fusion,
but their usage of  confidence is restricted to their
Web-link based models. We need a principled way that can incorporate confidence to
other types of models and can apply even when confidence assignments from different
extractors are of different qualities.

\subsubsection*{6. Using data from low-coverage sources judiciously}

Reliably computing the trustworthiness of data sources requires a sufficient
sample size, which is problematic when we start stratifying provenances
by extractor, pattern, etc.
As we have shown in Figure~\ref{fig:filter}, filtering out provenances with
low sample size can significantly improve probability prediction,
but this can result in triples without any probability estimate.
A more refined approach is called for, perhaps using hierarchical
Bayesian analysis to borrow statistical strength from related provenances.

\subsubsection*{7. Improving the closed world assumption}

We used the local closed world assumption
when we created the gold standard, which is used in
result evaluation and in initializing provenance accuracy in semi-supervised fusion.
However, in many cases this assumption is invalid,
either because {\em Freebase} does not know
all of the correct values for a data item with a non-functional predicate,
or because {\em Freebase} does not know the more specific values for a
data item with a functional predicate.
As we have seen in our error analysis,
50\% of the false positives are actually correct
decisions by our models but identified as wrong under the LCWA assumption.

A more creative way is needed to relax this assumption while still being able to
produce negatives. One possible solution is to associate a confidence with
each ground truth in the gold standard; the confidence can be associated with
the functionality of the predicated as we just described.
Such a gold standard with uncertainty allows us to give a lower penalty
for conflicts with uncertain ground truths.
(A related idea is discussed in~\cite{Ritter13}.)

\subsubsection*{8. Knowledge fusion for an open domain}
In this paper we focus on knowledge fusion where the subject
and predicate of each triple already exist in a knowledge base
such as {\em Freebase}. In the future, 
we also wish to enrich existing knowledge bases
by adding new entities, predicates, and types. Discovering new entities can
be considered as applying reconciliation on extracted triples;
discovering new predicates have been studied in~\cite{GHW+14}.
This process introduces even more types of noise that knowledge
fusion needs to be aware of, such as triples that are about the same new entity
but fail to be merged, new predicates that might be duplicates of existing ones,
and new predicates that are associated with wrong entity types for the subject and object.

\eat{
\subsubsection*{VII. Reasoning about dependent triples jointly}
Existing DF works assume independence between different data items.
For the knowledge that we have extracted, some of them are closely associated.
For example, for each spouse of {\em Tom Cruise}, we know the starting year
of the marriage, the ending year of the marriage, and sometimes the place of the wedding.
Other examples of dependent triples include combining a triple with its time stamp,
combining a numerical value with its unit, and so on.
In our data 39\% of the triples and 59\% of the data items fall in this case.
Such dependent triples call for joint reasoning about the truthfulness
of their components.
}

\section{Conclusions}
\label{sec:conclude}
In this paper we studied how far we can go in applying data fusion techniques to solve
the much harder problem of knowledge fusion. We carefully selected
and adapted three existing DF techniques, improved them with a series of refinements,
and analyzed their performance on the 1.6B unique knowledge triples 
we have extracted using 12 extractors from 1B+ data sources.
Although we show that the existing DF techniques work quite well for solving
the KF problem, we also discuss in detail their inherent limitations,
and we point out new directions 
for the research community to improve data fusion techniques to make
them more suitable for knowledge fusion.

\section{Acknowledgements}
We thank Fernando Pereira, Divesh Srivastava, and Amar Subramanya for 
constructive suggestions that helped us improve this paper. 
We also thank Anish Das Sarma, Alon Halevy, Kevin Lerman, Abhijit Mahabal,
and Oksana Yakhnenko for help with the extraction pipeline.

\balance

{\small
\bibliographystyle{abbrv}
\bibliography{base}
}

\end{document}